\documentclass[journal=jpcbfk,manuscript=article]{achemso}
\setkeys{acs}{usetitle = true}

\usepackage{color}

\author{R. Gonzalo Parra}
\author{Roc\'io Espada}
\author{Ignacio E. S\'anchez}
\affiliation[UBA-CONICET-IQUIBICEN]{Protein Physiology Lab, Dep de Qu\'imica Biol\'ogica, Facultad de Ciencias Exactas y Naturales, UBA-CONICET-IQUIBICEN, Buenos Aires, Argentina.}
\author{Manfred J. Sippl}
\affiliation[CAME-Univ Salzburg]{Center of Applied Molecular Engineering, Division of Bioinformatics, Department of Molecular Biology, University of Salzburg, Austria.}
\author{Diego U. Ferreiro}
\email{ferreiro@qb.fcen.uba.ar}
\phone{ (++54 +11) 4576-3300 }
\affiliation[UBA-CONICET-IQUIBICEN]{Protein Physiology Lab, Dep de Qu\'imica Biol\'ogica, Facultad de Ciencias Exactas y Naturales, UBA-CONICET-IQUIBICEN, Buenos Aires, Argentina.}

\title{Detecting Repetitions and Periodicities in Proteins by Tiling the Structural Space}

\begin{document}
\begin{abstract}

The notion of energy landscapes provides conceptual tools for understanding the complexities of protein folding and function. Energy Landscape Theory indicates that it is much easier to find sequences that satisfy the ``Principle of Minimal Frustration'' when the folded structure is symmetric (Wolynes, P. G. Symmetry and the Energy Landscapes of Biomolecules. Proc. Natl. Acad. Sci. U.S.A. 1996, 93, 14249-14255). Similarly, repeats and structural mosaics may be fundamentally related to landscapes with multiple embedded funnels. Here we present analytical tools to detect and compare structural repetitions in protein molecules. By an exhaustive analysis of the distribution of structural repeats using a robust metric we define those portions of a protein molecule that best describe the overall structure as a tessellation of basic units. The patterns produced by such tessellations provide intuitive representations of the repeating regions and their association towards higher order arrangements. We find that some protein architectures can be described as nearly periodic, while in others clear separations between repetitions exist. Since the method is independent of amino acid sequence information we can identify structural units that can be encoded by a variety of distinct amino acid sequences.

\end{abstract}



\small

\section*{Introduction}
 {\it
``There is something breathtaking about the basic forms of crystals. They are in no sense a discovery of the human mind; they just ``are'', existing quite independently of us. The most that man can do is become aware, in a moment of clarity, that they are there, and take cognizance of them.''} M.C. Escher

	Natural protein molecules are peculiar polymers. Unlike most of the random amino acid sequences, natural protein chains spontaneously find functional states by folding to a discrete collection of structures constituting a {\it native} state. Our deepest understanding of this phenomena is grounded in the Energy Landscape Theory of protein folding, which simplifies the complexity of folding to a few general descriptors of the configurational space \cite{pmid7784423,pmid16934172}. These abstractions provide conceptual tools to infer reliable energy functions \cite{pmid15664893} and to build simple and powerful predictive models \cite{pmid22545654,pmid23129648} and, most importantly, they provide a common language for the development (and healthy discussion!) of ideas \cite{pmid23112193,pmid16780604}. The basic notion underlying these developments is the {\it Principle of Minimal Frustration} \cite{pmid3478708}: in order to fold to a stable structure, a polymer must possess a funneled energy landscape.
	
	According to Energy Landscape Theory proteins are information-bearing molecules that evolved to funneled energy surfaces, contrasting them to random heteropolymeric chains that have rugged energy landscapes \cite{pmid7784423}. Since amino acids in natural proteins generally appear to be distributed at random\cite{pmid10988023}, higher order correlations must be present in sequences that result in stable folds. Energy Landscape Theory predicts that funneled landscapes and low energy structures are much easier to realize in the presence of symmetry as compared to asymmetric arrangements\cite{pmid8962034}. The identification of funneled energy landscapes as a general requirement for stable folds implies that patterns can form in different parts of the molecule with relative independence which subsequently assemble to higher order structures. This greatly reduces the search problem by efficiently arranging relatively small fundamental building blocks or ``foldons'' \cite{pmid8700876} in a repetitive fashion. The mere existence of repetitions or fundamental units does not guarantee that the system will be symmetric, but these units should arrange in particular ways and coalesce into higher order patterns. Hence a periodicity guarantees a certain symmetry but there can be repetitions without symmetry. Therefore, detecting repeated units and patterns is a first step towards an understanding of their assembly to complete structures and the emergence of symmetry. Such structural mosaics are accompanied by energy landscapes with multiple funnels embedded within each other \cite{pmidWalessym98,pmid18632565,pmid19368477}.

	Several algorithms have been used to characterize repetitions in protein sequences \cite{pmid23418055,pmid21884799}. Most methods are based on the self-alignment of the primary structure, while others implement spectral analysis of pseudochemical characteristics of the amino acids \cite{pmid23418055}. Since the same structural motif can be encoded by sequences that appear completely unrelated, it is not surprising that sequence-based methods fail to infer true structural repetitions when the sequence similarity is low. In contrast to sequence based methods, only a few methods for the detection of repetitions in protein structures are available. These usually search for repeats by aligning the structure against itself \cite{pmid15229884,pmid18487242}. Some methods add sophisticated transformations of the alignment matrices that enhance the detection and characterization of structural repeats \cite{pmid15340924,pmid11917144}, and machine learning provided a fast method to recognize repeat regions in solenoid structures \cite{pmid22962341}. Although many families of proteins with repeating motifs can be identified \cite{pmid10512723,pmid21884799}, there is still no consensus on how to reconcile the often conflicting characterizations of repetitions in proteins \cite{pmid22923522,pmid23418055} even for basic parameters such as the size of the repeating elements, the number and location of the occurrences and the grouping of these into higher order patterns.
	
	Here we develop basic concepts and methods for the detection and analysis of repeats in protein structures. Using a fast and robust structural alignment protocol and a proper metric \cite{pmid22483118}, we exhaustively analyze the repetition of every possible continuous fragment of a protein structure and define the portions that best describe the overall structure when this fragment is repeated, translated and rotated exhaustively with respect to the complete molecule. The result is a tessellation of the whole protein in terms of a set of basic tiles. The tessellation lends itself to an intuitive visualization of the repeating units and their association into higher order patterns. We find that some architectures can be described as nearly periodic, while in some others clear separations between repetitions exist. Since this method is independent of sequence it allows for comparison of recurring structures and tiles that represent a common structural motif that can be encoded by a variety of distinct sequence elements.

\section*{Methods}

\subsection*{Structural alignments and tiles}

For the characterization of repetitions and the identification of tiles in protein structures we use the TopMatch tool \cite{pmid18174182,pmid22483118}. Given a pair of protein structures this algorithm generates an exhaustive list of partial alignments along with the transformations (rotations and translations) that maximize the superposition of equivalent C$^\alpha$ atoms. The alignments are ranked according to the TopMatch score,
\[
S=\sum_i^L e^{-r_i^2/\sigma^2}
\]
which provides a metric for structural similarity \cite{pmid18227113}. Here $L$ is the length of the alignment and $r_i$ is the euclidean distance between equivalent C$^\alpha$ atoms. Basically $S$ is a function of the alignment length $L$ and the structural deviation of the superimposed structural fragments, where the scaling factor $\sigma$ determines the rate of reduction of $L$ as a function of the structural deviation. Here we used $\sigma=6.35$ {\AA} as reported previously\cite{pmid22483118}.
Proteins often contain recurrent structural motifs that can be considered as repetitions and variations of a basic structural unit. In order to detect this kind of structural repetition, we treat the structure as a mosaic and try to decompose it into smaller units or {\em tiles} with the constraint that these tiles are all structurally similar to each other. In a protein the possible tiles are not necessarily unique nor are they required to cover a chain completely. But in any case, it is certainly possible to identify those tiles that, when repeated in a non-overlapping fashion, cover a maximum fraction of the structure.

Given a protein structure, every continuous fragment of the polypeptide is a possible tile. Hence the length of tiles ranges from the sequence length $N$ down to a single residue. Since the C$^\alpha$ traces of tiles of one or a few residues are too small for meaningful comparisons we use a lower tile length of six amino acid residues. In a protein of length $N$ there is one tile of length $N$, two tiles of length $N-1$, and so on, and hence the total number of tiles is $N_T = \sum_{L=6}^{N}(N-L+1)$. Each of these tiles $T_i$ is then used as a query in TopMatch to identify all other tiles $T_k$ that are structurally similar to $T_i$. Each match is uniquely identified by its length $L_{ik}$, the location of its center $Z_{ik}$, and the associated score $S_{ik}$. The matches are then sorted by $S_{ik}$, where the self-alignment ($i \equiv k$) necessarily has the highest score since the respective alignment length is maximal and the structural deviation is zero. Hence $L_{ii} = S_{ii}$, i.e. the score obtained from an alignment of a tile with itself evaluates to the length $L_{ii}$ of the alignment.

From the set of matches we extract that subset of fragments that maximizes the sum over the scores $C_i = \max{\sum_k S_{ik}}$, where any two tiles $T_{k_1}$ and $T_{k_2}$ that occur in the sum must not overlap. This sum defines the coverage $C_i$ of tile $T_i$ which was used to generate the matches. We define the associated tile score as

\begin{equation}
	\Theta_i = \frac{C_i-L_{ii}}{N-L_{ii}}
	\label{eq:ts}
\end{equation}
which represents the fraction of the structural space that can be covered by repetitions of a given tile.
When considering the ranked list of hits there are several ways to define the set of non overlapping alignments. In the most restrictive variant we include only those repeats $T_k$ for which the aligned region comprises the whole tile, i.e. $L_{ik} \equiv L_{ii}$. A more flexible variant is to include all alignments where $L_{ii}/2 < L_{ik} \leq L_{ii}$, that is when more than half of $T_k$ matches $T_i$. In the latter case we use the additional restriction that the first and last residues of any two tiles $T_h$ and $T_k$ in the optimal subset must not overlap.

\subsection*{Homogeneous model}

To evaluate the upper limits of the tiling scoring functions we calculated the tile score $\Theta_i$ expected for a homogenous model, where the protein is represented as a finite linear string of amino acids. In this case, every alignment of tile $T_i$ and repeat $T_k$ has a perfect match, and thus the alignment score $S_{ik}$ will be equal to $L_{ii}$. Then the coverage $C_i$ is  the product of $L_{ii}$ and the number of tile copies $n_c$ that can be accommodated which, depending on the tile center $Z_i$, is $n_c = \lfloor\frac{N}{L_i}\rfloor$ if the chain ends are covered or $n_c = \lfloor\frac{N}{L_i}\rfloor -1 $ if they are not.

When alignments with $L_{ii}/2 < L_{ik} \leq L_{ii}$ are permitted, then $\Theta_i$ has an extra term that takes into account the coverage at the chain ends:
\begin{equation}
\Theta_i=\frac{(n_c-1)\cdot L_{ii}+C_{beg}\cdot \chi \left(C_{beg}-L_{ii}/2\right)+C_{end}\cdot \chi \left(C_{end}-L_{ii}/2\right)}{N-L_{ii}}
\label{eq:TS-PSM}
\end{equation}
where $\chi (x) = \left\lbrace \begin{array}{ccc}
0 & \mbox{if} &x < 0\\
1 & \mbox{if} & x \geq 0
\end{array} \right.$, $n_c$ is the number of full length tile copies that can be accommodated along the protein, and $C_{beg}$ and $C_{end}$ are the maximum number of amino acids left uncovered by the copies at the limits of the protein, and can be calculated as:
\begin{eqnarray}
C_{beg}&=& Z_i+\lceil\frac{1}{2}-\frac{Z_i}{L_{i}}\rceil \cdot L_{i}-\frac{L_{i}}{2}\\
C_{end}&=& N-\left[Z_i+\lfloor\frac{N}{L_{i}}-\frac{1}{2}-\frac{Z_i}{L_{i}}\rfloor\cdot L_{i}+\frac{L_{i}}{2}\right]
\end{eqnarray}
Further details of this model can be found in the supplementary material.

\section*{Results and Discussion}

\par To illustrate the characteristic properties of tessellations of protein structures we use the protein '4ank' (pdb:1n0r, $N=126$ residues) which is a synthetic construct of canonical ankyrin repeats\cite{pmid12461176}. Figure 1a shows the scores of the top 15 hits for 3 different fragments of the structure used as the query tile $T_i$. In all instances the highest ranking tile corresponds to the self-alignment ($i \equiv k$) and in each of these cases there are two tiles ($i\neq k$) that yield nearly perfect matches. For the subsequent tiles the score drops rapidly.

\medskip
	\begin{figure}
\centering
	\includegraphics[width=0.5\textwidth]{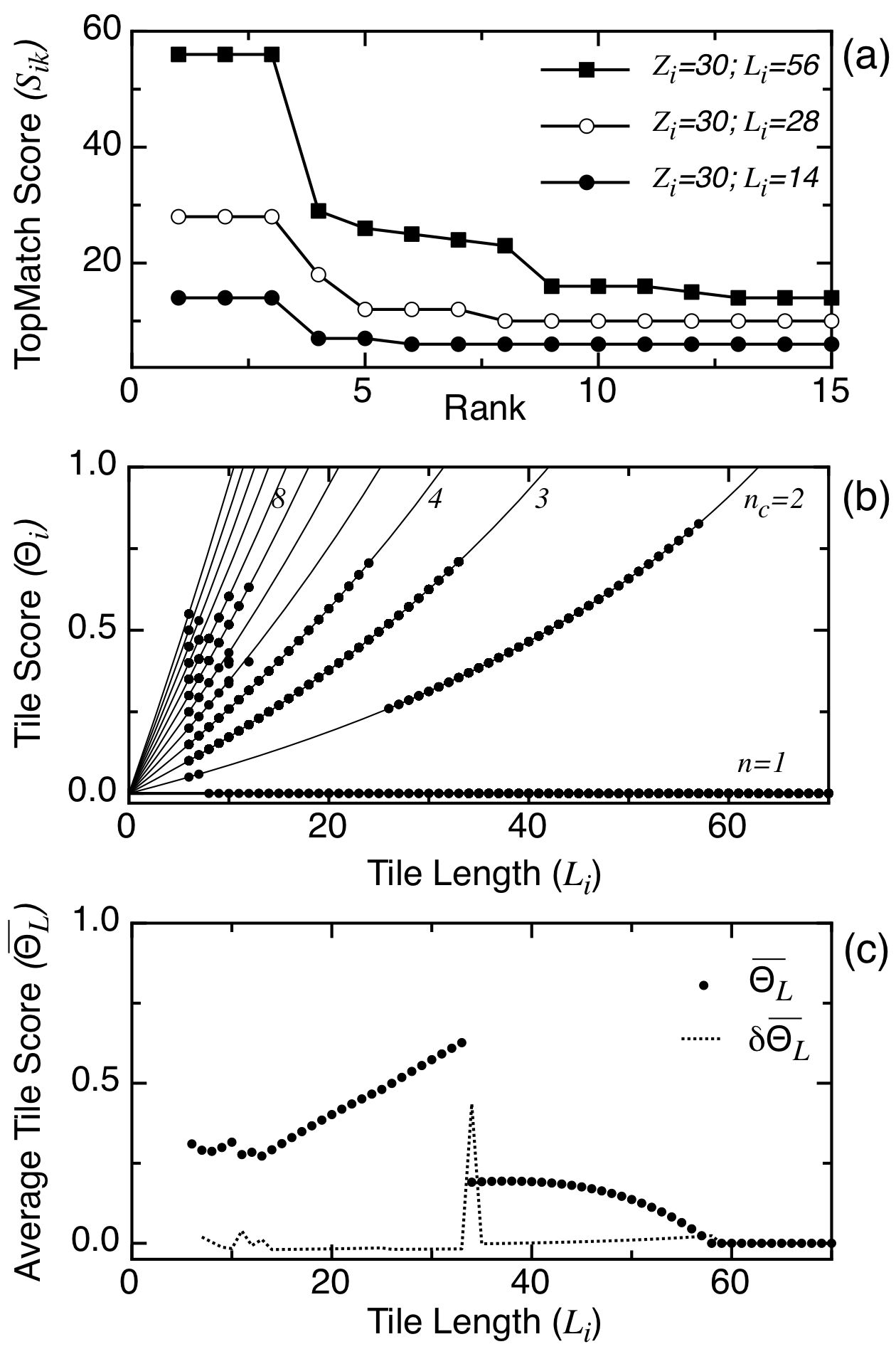}
	\caption{ Scoring the Tiles: continuous portions of a protein model (pdb:1n0r,A) are selected and structurally aligned to the whole protein. A ranked list of alignments is generated for every fragment according to TopMatch score ($S_{ik}$), three of which are shown in panel a). $L_{i}$ is the length of the fragment in amino acid units and $Z_{i}$ is the center, according to the numbering scheme of the C$^\alpha$ atoms of the pdb. b) Distributions of tile score $\Theta_i$ for every tile length ($L_{i}$). Each point corresponds to the experimental values obtained when perfect matching ($S_{ik}=L_{ii}$) is restricted. The lines correspond to the expression $\Theta_i=(n_c -1)L_{ii}/(N-L_{ii})$ with $N=126$ and the number of tile copies that can be repeated is $n_c=1,2,3 ...12$ as indicated. c) The points correspond to the average $\overline{\Theta_i}$ calculated for every $L_{i}$. The dotted line is the difference between consecutive points $\delta\overline{\Theta_i}$.
	}
	\label{fig:crude}
\end{figure}
\medskip

Next we use the ranked list to pick out non-overlapping fragments in order to cover the protein structure as repeats of tile $T_i$. For each possible tile $T_i$ the tile score $\Theta_i$ is calculated as described above.  Clearly, the tile score $\Theta_i$ for tiles with $L_{ii} > N/2$ is always zero, as no repetitions of such fragments are possible (Fig 1b). The largest tile that can be repeated twice has $L_{i}=57$ amino acids. Tiles nested within these largest tiles necessarily have smaller scores. Three repeats are observed for $L_{i}=33$, and four for $L_{i}=24$ (Fig 1b). The peaks in Figure 1b correspond to the largest fragments that occur more than once and for which each of its extensions occurs fewer times, that is, they are {\it maximal elements}. The steady decrease in $\Theta_i$ results from fragments that are nested within the maximal ones. This can be inferred from the homogenous model where a group of tiles that occurs $n_c$ times yields the tile score $\Theta_i=(n_c -1)L_{i}/(N-L_{i})$.

The fact that there is a number of tiles of similar score $\Theta_i$ but varying length $L_i$ implies that the overall protein architecture can be covered by a set of nested tiles. Hence, the question arises which of the possible tile lengths yields a tessellation of maximum coverage. In the case of real proteins, copies of individual tiles generally exhibit structural variations with respect to a basic tile. Such variations reduce the score $S_{ik}$ of the respective structural matches. The relative reduction is generally much more pronounced for small tiles as compared to larger tiles which may result in a relatively large decrease of the overall tile score $\Theta_i$. In short, if the various copies of small tiles have relatively large structural deviations then the associated tile score $\Theta_i$ may appear suboptimal with respect to tile scores obtained from larger tiles.
It is therefore convenient to take the average $\overline{\Theta_L}$ over all tile scores $\Theta_i$ that have the same tile length $L_{i}$ (Fig 1C). In the example it is evident that the maximum occurs at $\overline{\Theta}_L=33$ residues, indicating that tiles of this size tessellate the structure in an optimal way. Formally, the optimal length  is obtained as a root of the derivative $d\overline{\Theta}_L/dL$, i.e. it can be obtained from the finite differences  $\Delta\overline{\Theta}_L=\overline{\Theta}_L-\overline{\Theta}_{L-1}$. Note that this identifies the optimal tile length $L$, but not the particular tile $T_i$ that optimizes the tessellation.

Since a particular tile $T_i$ is characterized by the position of its center $Z_i$ along the amino acid sequence and a match between two tiles $T_i$ and $T_k$ by the respective alignment length $L_{ik}$ the multitude of tessellations of a particular structure is representable in two dimensions and the associated score $\Theta_i(L_i,Z_i)$ can be indicated by shades of gray (Figure 2). Such representations show how copies of each of the possible tiles cover the whole structure. In the case of 1n0r, the structure is covered by two repeats of 57 amino acids, centered at residues 30 and 96. These repeats decay into two smaller repeats of 24 amino acids, where the decomposition results in a loss of approximately 12\% coverage. These tiles in turn consist of two smaller tiles of 8 and 10 amino acids. The latter correspond to two $\alpha$-helices that are part of the canonical ankyrin motif (Figure 2).

A peculiar phenomenon is apparent for tiles of length $L_i=33$. Any tile of this size provides a nearly complete tessellation of the structure. Moreover, at this length scale the tiles are separated by a distance that is equal to the size of the tile itself. Hence, the whole structure has the characteristics of a wave. The characteristic wave length is $L = 33$, and the structure can be completely covered starting with any phase $\phi = 0,1,\ldots,L-1$.
Taken together these observations imply that those tiles optimally cover a repetitive protein structure whose average score $\overline{\Theta_L}$ is a maximum and it seems that such maxima are accompanied by a large value of $\Delta\overline{\Theta}_L$ (Figure 1b). From the set of tiles that contribute to $\overline{\Theta_L}$ we may define the most typical tile as that particular tile $T_i$ that has the largest score $\Theta_i(L_i,Z_i)$ with respect to all other tiles $T_k(L_i)$ in this set.

Repeats in protein structures are thought to be the result of duplication of amino acid sequences. In general a duplication results in an exact copy of the duplicated material. On the level of amino acid sequences the similarity among the copies decays in time due to the accumulation of amino acid substitutions, insertions, and deletions. The respective structures are more robust in the sense that the similarity among the sequences decays much faster as compared to the similarity among the polypeptide backbone. Nevertheless, insertions, deletions and other events also affect the three dimensional structures of the individual copies and therefore, in natural proteins structural repeats are rarely exact and they are often interspersed by non-repetitive regions. In what follows we discuss tessellations obtained for a broad variety of protein structures. This method does not rely on visual inspection. We define the characteristic frequency at the highest peak in $\Delta\overline{\Theta}_L$, and the basic tile-unit as the one that scores highest $\Theta_i$ at this $L_{i}$. The non-repetitive regions found in these tessellations are marked as insertions (Table S1).

\medskip
	\begin{figure}
\centering
	\includegraphics[width=0.9\textwidth]{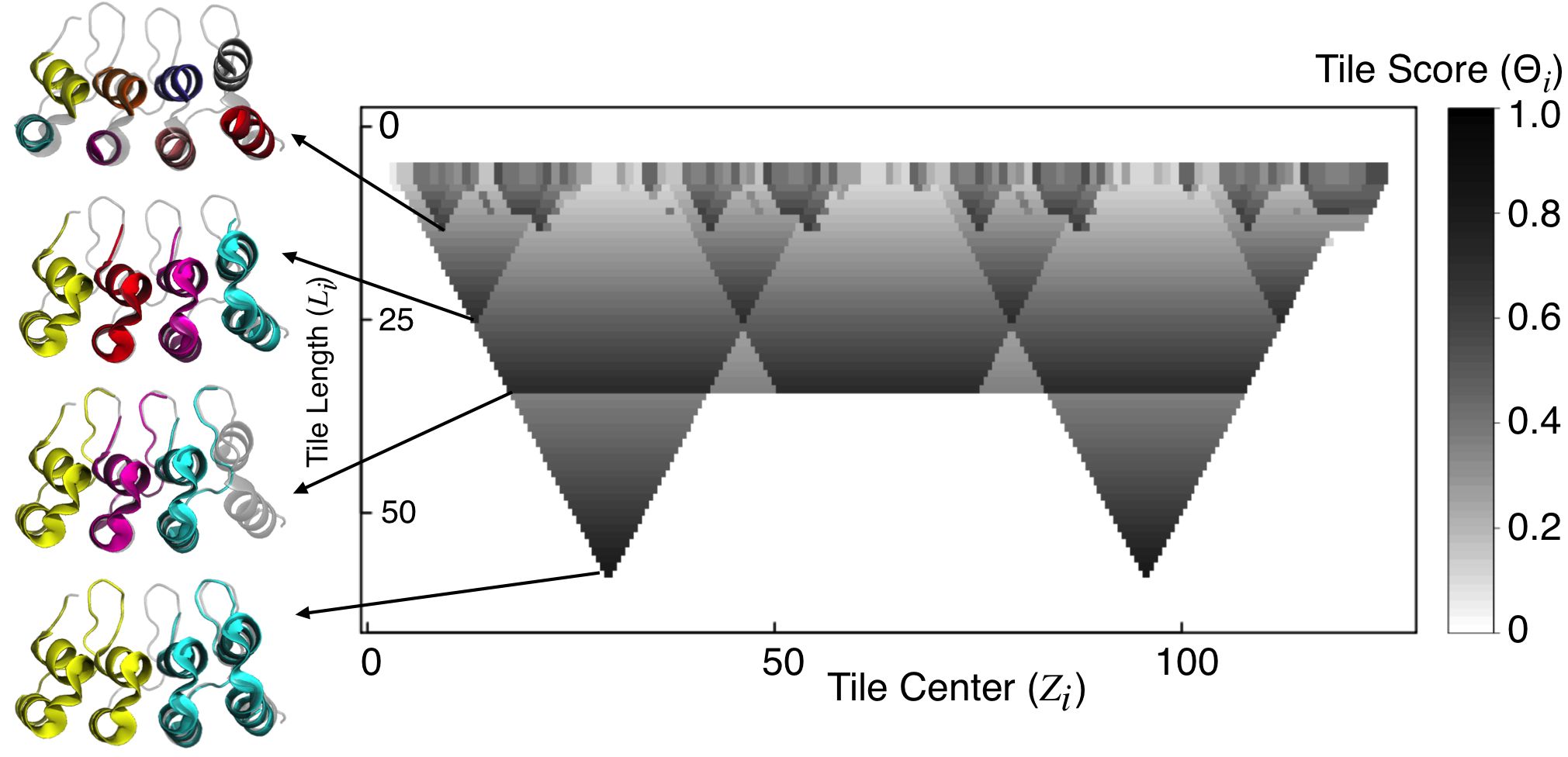}
	\caption{ Tiling a highly symmetric protein: a designed ankyrin-repeat protein (pdb: 1n0r,A) was fragmented in 7381 different tiles. These are ordered according to their size (vertical axis) and their center (horizontal axis) in amino acid units. The tile score $\Theta_i$ of each one is displayed in grayscale. The structures of the protein and the respective tiling at different $L_{i}$ is shown on the left. The native structure is colored gray, and superimposed to it is the selected tile (yellow), and the copies of it colored cyan, magenta, red, etc. }	
	\label{fig:1n0ra}
\end{figure}
\medskip

\subsection*{Tessellations of classical repetitive proteins}
	Many natural proteins contain tandem repeats of similar amino acid stretches. They are broadly classified in groups according to the length of the minimal repeating unit. Short repeats of up to five residues usually form fibrillar structures such as collagen or silk, while repeats longer than about 100 residues frequently fold independently as globular domains \cite{pmid21884799,pmid9177189}. There is a class of repeat proteins that lies in between these for which folding of the repeating units is coupled and ``domains'' are not obvious to define \cite{pmid18483553}. Since defects in the regularities of the repeating array are likely to affect the folding transitions and the biological function, we aimed at defining these from a purely geometrical perspective using the tiling approach described above.
	

I$\kappa$B$\alpha$ is an ankyrin repeat containing protein that binds to and inhibits the transcription factor NF-$\kappa$B \cite{pmid20055496}. The fragmentation and tiling procedure correctly identifies a characteristic 33 amino acids length corresponding to the canonical ankyrin repeat size (Figure 3a). We found deviations from this canonical size ranging from 30 to 39 residues, indicating that not all the ankyrin repeats are geometrically equivalent. Fragments with highest scores can be placed 6 times, covering about 92\% of the structure (Table S1). It is apparent that the most C-terminal repetition is distorted relative to the others, as the $\Theta_i$ corresponding to this region are lower. The grouping of consecutive repeats at bigger $L_{i}$ segregate pairs where the central one scores best, indicating that the insertions detected at length 33 distort the symmetry of the array at a higher length scale. Maybe it is no coincidence that this protein was shown to fold {\it in vitro} in three consecutive transitions roughly corresponding with the pairing of repeats at $L_i=70$ \cite{pmid21329696,pmid17174335} .

\medskip
	\begin{figure}
\centering
	\includegraphics[width=0.9\textwidth]{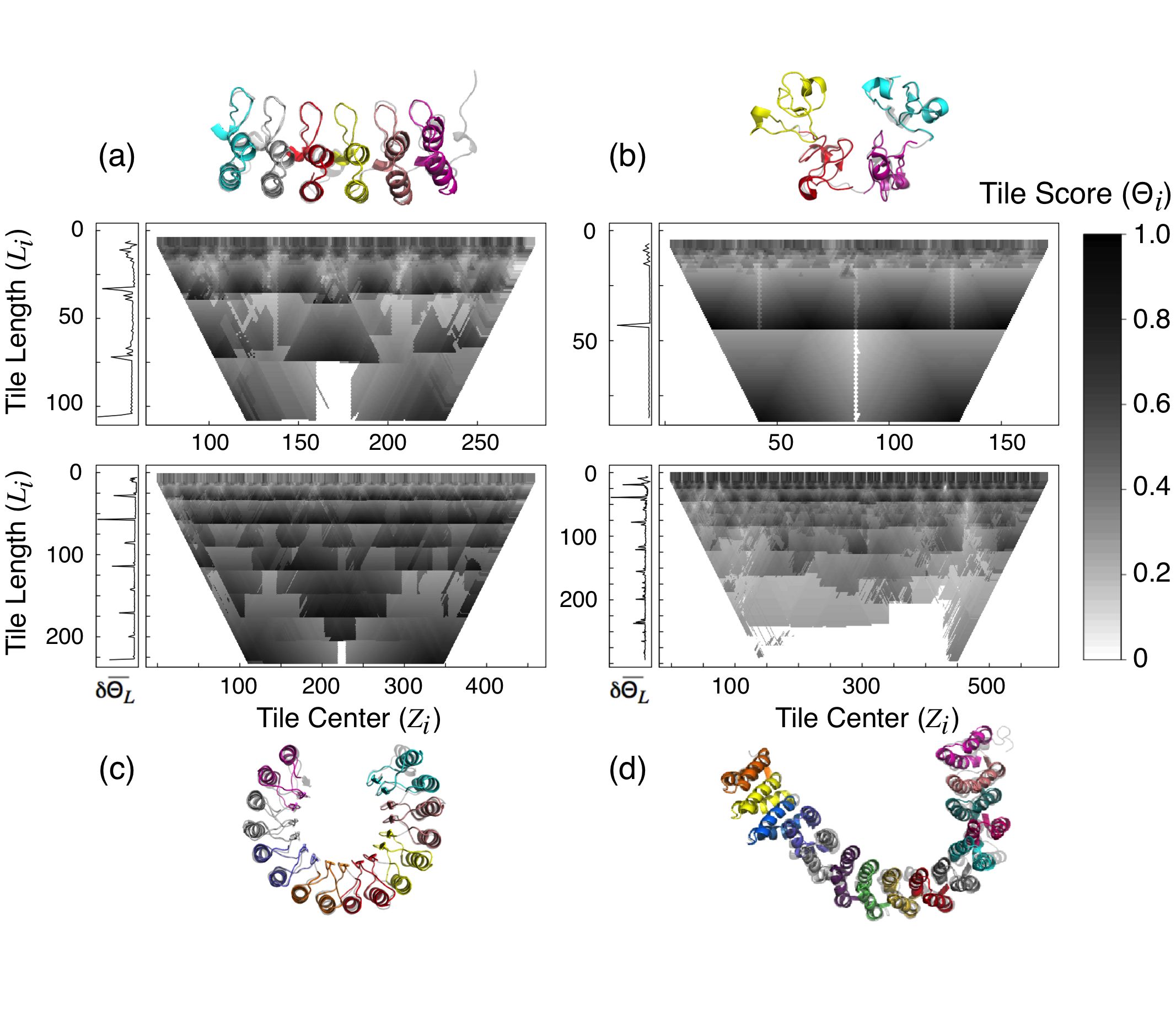}
	\caption{ Tiling classical repeat-containing proteins. The tiling profile is shown on grayscale, together with the $\delta\overline{\Theta_L}$ projected on the left. The structures of the native protein and the highest scoring tiling at the characteristic frequency are shown, using the same coloring scheme of Fig.2. The length ($L_{i}$) and center ($Z_{i}$) of the selected tile is: a) Ankyrin repeat: I$\kappa$B$\alpha$ (pdb:1nfi,E) $L_i=33, Z_{i}=191.5$ b) Hevein: wheat-germ agglutinin (pdb:1k7u,A) $L_i=43, Z_{i}=150.5$ c) Leucine-rich: Porcine ribonuclease inhibitor (pdb:2bnh,A) $L_i=57, Z_{i}=139.5$ d) HEAT: PR65/A (pdb:1b3u,A) $L_i=39, Z_{i}=530.5$ }		
	\label{fig:classrepeats}
\end{figure}
\medskip


The monomeric chain of wheat-germ agglutinin has been described to contain four hevein subdomains \cite{pmid11853952}. The tiling approach detects that this structure can be composed with 2 tiles of $L_i=86$ amino acids, as well as 4 repetitions of $L_i=43$, both covering 100\% of the structure (Fig 3b). Taking the average of the $\Theta_i$ at each $L_{i}$ points that a discontinuity occurs at size 43, defining a characteristic frequency. At this size most tiles are equally good in covering the structural space with three repetitions. The highly symmetric disposition of the four best tiles at this length scale makes the whole structure appear nearly periodic, and a preferred phase is determined by the N and C termini of the chain.

Porcine ribonuclease inhibitor is a leucine-rich repeat protein for which 16 consecutive repetitions were defined in its sequence. Although very similar at the primary level, these repeats are not structurally equivalent. We detect that there are two different types of tiles, each consisting of 28 and 29 amino acids (Fig 3c). Moreover, we found that these are alternated along the structure, appearing as a square-tooth pattern at this length scale (Fig S3). Since these units are arranged in a symmetric fashion, the structure can be represented as well by bigger fragments (Fig 3c). At the length of $L_i=57$ residues, almost every fragment repetition is as good as others in explaining the overall structure. Thus, the repeating length is better described with two canonical leucine-rich repeats. It is striking to note that Haigis {\it et al} previously identified a 57 residue repeat as the evolutionary unit of this protein by analyzing the exon boundaries of the primary transcripts \cite{pmid12032252}.

The scaffolding subunit of protein phosphatase 2A, PR65/A, is a large repeat-protein of the HEAT class \cite{pmid9989501}. The tiling procedure detects the best tile at size 39 amino acids and identifies 15 copies of it in the structure, coincident with the detection in amino acid sequence patterns of the HEAT motif (Fig 3d). This protein exhibits an overall superhelical structure, yet irregularities in the array cause unevenness in the grouping of consecutive repeats at higher length scales. The periodic packing of HEAT repeats is interrupted between repeats 3 and 4 ($Z_i = 117$) and between 12 and 13 ($Z_i = 471$) \cite{pmid9989501}. This is reflected at higher $L_{i}$ where the tiles centered around amino acid 300 display consistent higher scores, indicating that the central repeats are more symmetrically arranged than the terminal ones (Fig 3d).

\subsection*{Tessellations of globular proteins}

	In contrast to the solenoidal architectures usually acquired by classical repeat-proteins, some protein folds display point rotational symmetries. Often the N and C terminal repetitions come in contact, closing up the structure in polyhedral-like forms. We investigated how the tiling procedure identifies structural repetitions and tessellation patterns in some of the most common topologies of this kind.

The TIM barrel is one of the most common folds among monomeric enzymes \cite{pmid12206759}. This is typically described as a collection of $\beta$-$\alpha$ motifs linked by variable loops that close up a cylinder of parallel $\beta$-strands surrounded by a layer of $\alpha$-helices. There is a relatively high structural conservation among proteins of this type, yet their sequences can appear unrelated, opening room for discussion about the nature of the repeating units and their arrangement \cite{pmid16844977}. We applied the tiling procedure on some of the most discussed cases and for most we detect signals for 2, 4 and 8 repeats (Table S1, Fig S4). Not all the TIM barrels showed the same characteristic frequency. Some of the structures are best described with fragments that correspond to half barrel (Fig 4a), while others displayed comparable signals at sizes corresponding to half or quarter barrel (Fig S4). The most irregular examples have characteristic frequencies at even lower length scales (Table S1). Based on amino acid sequence patterns, Soding {\it et al} annotated equivalent deviations in this topological family \cite{pmid16844977}.

Several proteins can be grouped into the $\beta$-propeller class. These contain a variable number of radially arranged antiparallel $\beta$-sheets appropriately named ``blades'' \cite{pmid10607670}. We identify that in most cases the best tiles distinguish this motif and annotate 4, 5, 6, and 7-bladed propellers (Fig. S5), even when a non-propeller domain is present in the same polypeptide chain (Fig S5d).
An interesting exception occurs in the subclass of WD-repeat propellers where the selected tile does not correspond with a blade (Fig 4b). In this case we detect a characteristic frequency of $L_i=42$ amino acids, with tiles repeated 7 times and contributing three strands to one blade and one strand to the next one (Fig 4b). Notably, this particular phase was the one originally described when no structure of members of this class were known \cite{pmid8090199}.

\medskip
	\begin{figure}
\centering
	\includegraphics[width=0.9\textwidth]{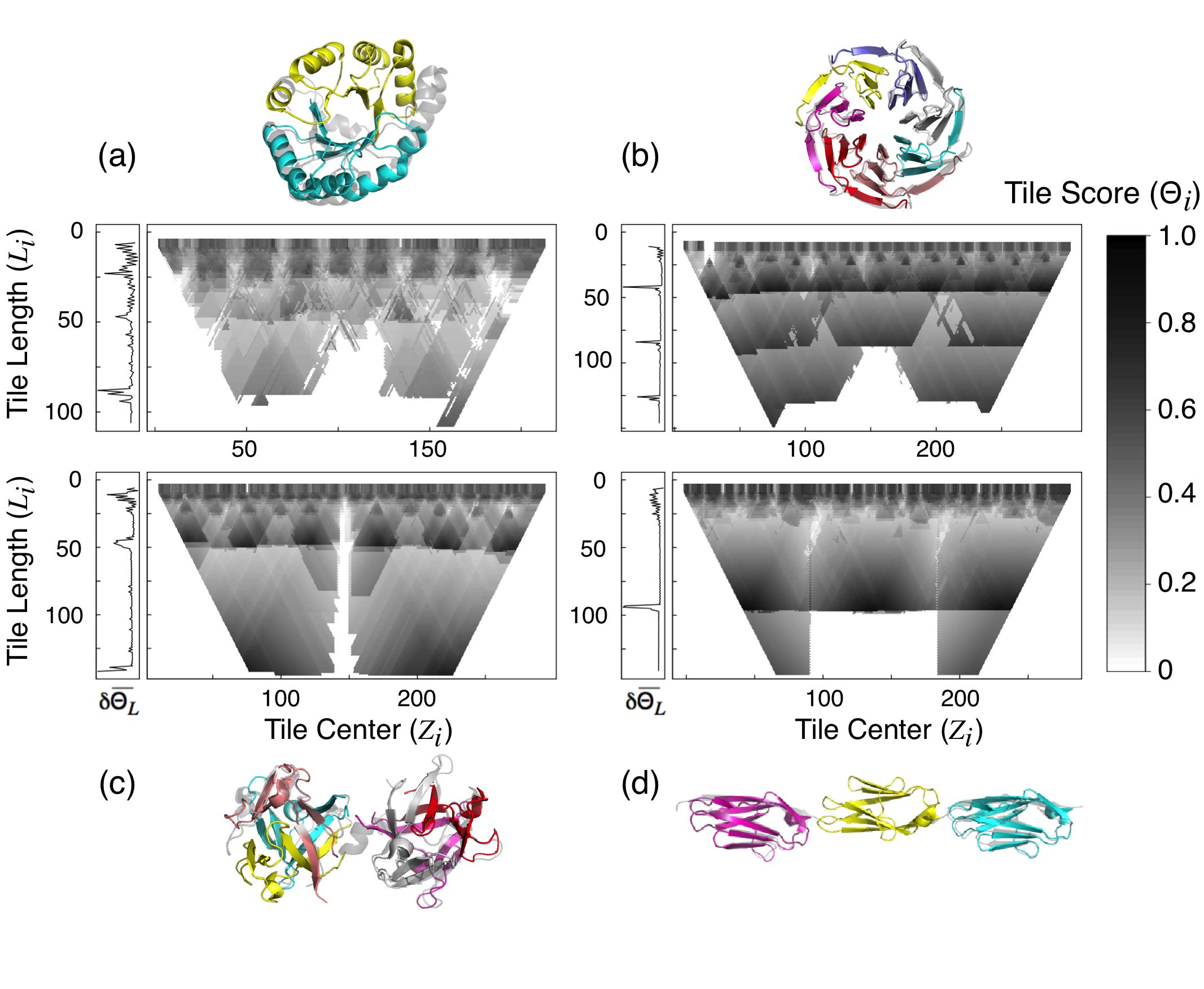}
	\caption{ Tiling globular repeat-containing topologies. The tiling profile is shown on grayscale, together with the $\delta\overline{\Theta_L}$ projected on the left. The structures of the native protein and the highest scoring tiling at the characteristic frequency are shown, using the same coloring scheme of Fig.2. The length ($L_{i}$) and center ($Z_{i}$) of the selected tile is: a) TIM barrel: (pdb:1fq0,A)  $L_i=88, Z_{i}=60$ b) $\beta$-propeller (pdb:3ow8,A)  $L_i=42, Z_{i}=200$  c) Trefoil (pdb:1ybi,A)  $L_i=142, Z_{i}=223$ d) Ig-repeats (pdb:2rik,A)  $L_i=94, Z_{i}=140$ }
	\label{fig:otherrepeats}
\end{figure}
\medskip

The hemagglutinating protein HA33 from {\it Clostridium botulinum} is a neurotoxin-associated protein that folds in an appealing topology of two  consecutive $\beta$-trefoil subdomains (Fig 4c). The characteristic frequency ($L_i=142$) points to two fragments that have the highest $\Theta_i$ and correspond to the tiles of each subdomain. The best phase at the second peak ($L_i=46$) correspond to tiles that can be fitted 3 times in each subdomain and match the annotated foil of the $\beta$-trefoil architecture.

Surveying other architectures with repeating motifs we noted that in some cases the highest scoring tiles are at the characteristic frequency. Figure 4d shows the results for a fragment of titin that contains 3 tandem immunoglobulin-like (Ig) domains. At $L_i=94$ amino acids, the best phase coincides with the Ig domains. The fact that other phases also score high at this length scale is indicative that the arrangement between the Ig domains is regular, as if this were not the case, those fragments would not display that high $\Theta_i$.

	At some level, all proteins are formed by repetitions of amino acids. The symmetry of the backbone interactions in secondary structures was key to Pauling and Corey proposal of these arising from the regular repetition of planar peptide bonds \cite{pmid14816373,pmid14834147}. Recurrent secondary structure motifs were once candidates for fundamental building blocks of globular domains, in line with the success of structure prediction by fragment assembly \cite{pmid21997831,pmid19706384,pmid9149153}. Since repetitions can be confidently found by tiling the structural space, we explored to what extent any given protein structure can be said to be composed with tiles, illustrating with some classical examples.

	Synthesized at embryonic stages and hopefully lasting soluble for a lifetime \cite{pmid21272671} $\beta\gamma$-crystallins lens proteins increase the refractory index and maintain transparency of the vertebrates' eyes. Since its initial description it has been a clear example of structural motifs coalescing into higher order patterns. Coincident with the classical descriptions of these fold, tiling the structural space detects that this protein can be very well described with two repetitions of an eight-stranded $\beta$-barrel of $L_i=87$ amino acids centered at position $Z_i=44.5$ and $Z_i=133.5$ (Fig 5a). In turn, each of these can be composed with two units of about 40 residues that correspond to the Greek-key motif, that can be further decomposed into three 10-residue $\beta$-strands. 
	The characteristic frequency is at $L_{i}=43$ amino acids, selecting out the greek-key as the repetition we annotate. It is apparent that there are irregularities in the structure that make the second and fourth greek-keys have a higher $\Theta_i$ than the others and indeed different maximal $L_{i}$.
	
About 70\% of the mean structure of Myoglobin, the hydrogen atom of biology \cite{pmid12861080}, can be described with 6 copies of an 18 amino acid fragment. This corresponds to `B' $\alpha$-helix, and constitutes a maximal fragment. The score at higher length scales decreases rapidly (Fig 5b). In this case we could not detect a relevant frequency above the $\alpha$-helical segments, indicating that these do not contiguously repeat in a highly symmetrical way, a fact that strongly surprised Kendrew {\it et al} when they solved the crystal structure \cite{pmid13517261}.

Green fluorescent protein folds as a $\beta$-barrel with a coaxial helix, with the fluorophore forming from the central helix \cite{pmid8703075}. We identify fragments of $L_i=15$ that can cover about 71\% of the structural space with 11 repetitions, corresponding with $\beta$-strands (Fig 5c). At higher length scales no fragment significantly raises the signal.

 {\it Bacillus licheniformis}  $\beta$-lactamase illustrates an example of a mixed $\alpha\beta$ topology, composed of two discontiguous subdomains \cite{pmid14769049}. Here again there is no particular length scale at which a useful characteristic frequency can be defined (Fig 5d). The best tiling occurs at $\Theta_i=15$ where the fragment corresponds to one of 10 $\alpha$-helices and covers 74\% of the structural space when repeated.

\medskip
	\begin{figure}
\centering
	\includegraphics[width=0.9\textwidth]{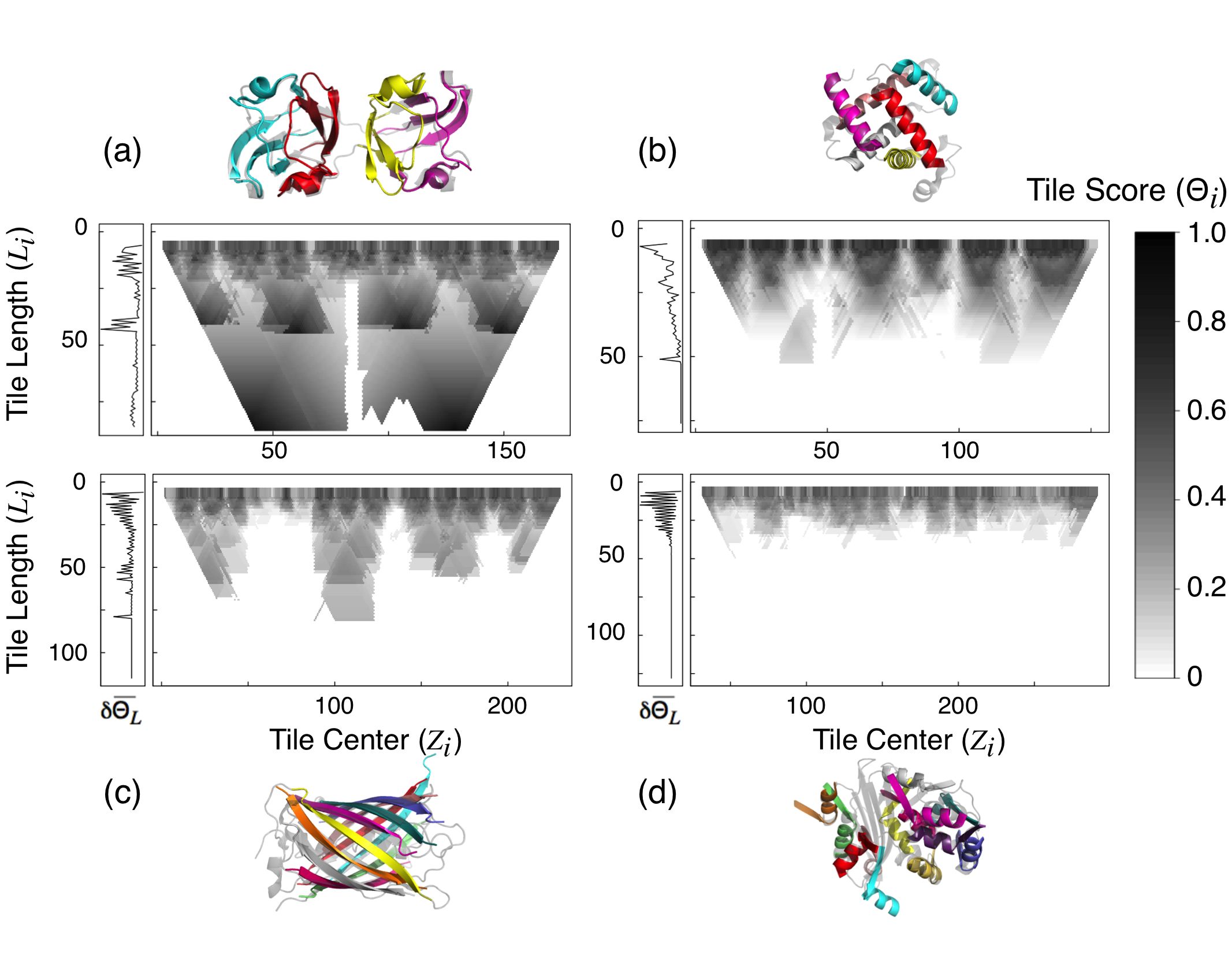}
	\caption{ Tiling classical globular proteins. The tiling profile is shown on grayscale, together with the $\delta\overline{\Theta_L}$ projected on the left. The structures of the native protein and example tilings are shown, using the same coloring scheme of Fig.2. The length ($L_{i}$) and center ($Z_{i}$) of the selected tile is: a) $\beta\gamma$-crystallin (pdb:1h4a,X)  $L_i=43, Z_{i}=149.5$ b) Myoglobin (pdb:1mbd,A)  $L_i=18, Z_{i}=29$ c) Green Fluorescent Protein (pdb:1gfl,A)  $L_i=15, Z_{i}=182.5$ d)  $\beta$-Lactamase(pdb:4blm,A)  $L_i=15, Z_{i}=185.5$ }
	\label{fig:norepeats}
\end{figure}
\medskip

\subsection*{Tessellations of oligomers}

In their natural environment, most of the polypeptide chains of living organisms are not found folded as spheroidal monomers, but typically come together forming oligomeric complexes with two or more subunits. Most frequently they form homo-dimeric complexes, but hetero-oligomers are not uncommon and even thousand-mers are to be found. The symmetrical basis of this phenomena have been explored even before the first protein structures were solved \cite{pmid10940245}. A recent survey estimates that over 95\% of the homodimeric complexes crystallized are symmetric \cite{pmid22629324}, and it is expected that small insertions and deletions can have profound effects on protein functionality, modulating oligomer stability, specificity and aggregation \cite{pmid21048085}.
To analyse the details of symmetry in multi-chains complexes we can first define the elementary blocks that constitute the array. To explore this we applied the same procedure of fragmenting and tiling described above now using the quaternary arrangements of subunits as the target structure to cover. If the monomers that form an homo-oligomer cannot be decomposed into significant tiles, we expect the best tile to correspond to the monomeric chain itself. Indeed we find this is the case for the majority of the oligomers we evaluated. We noted however interesting cases in which the subunits can be decomposed into significant tiles.

\medskip
	\begin{figure}
\centering
	\includegraphics[width=0.9\textwidth]{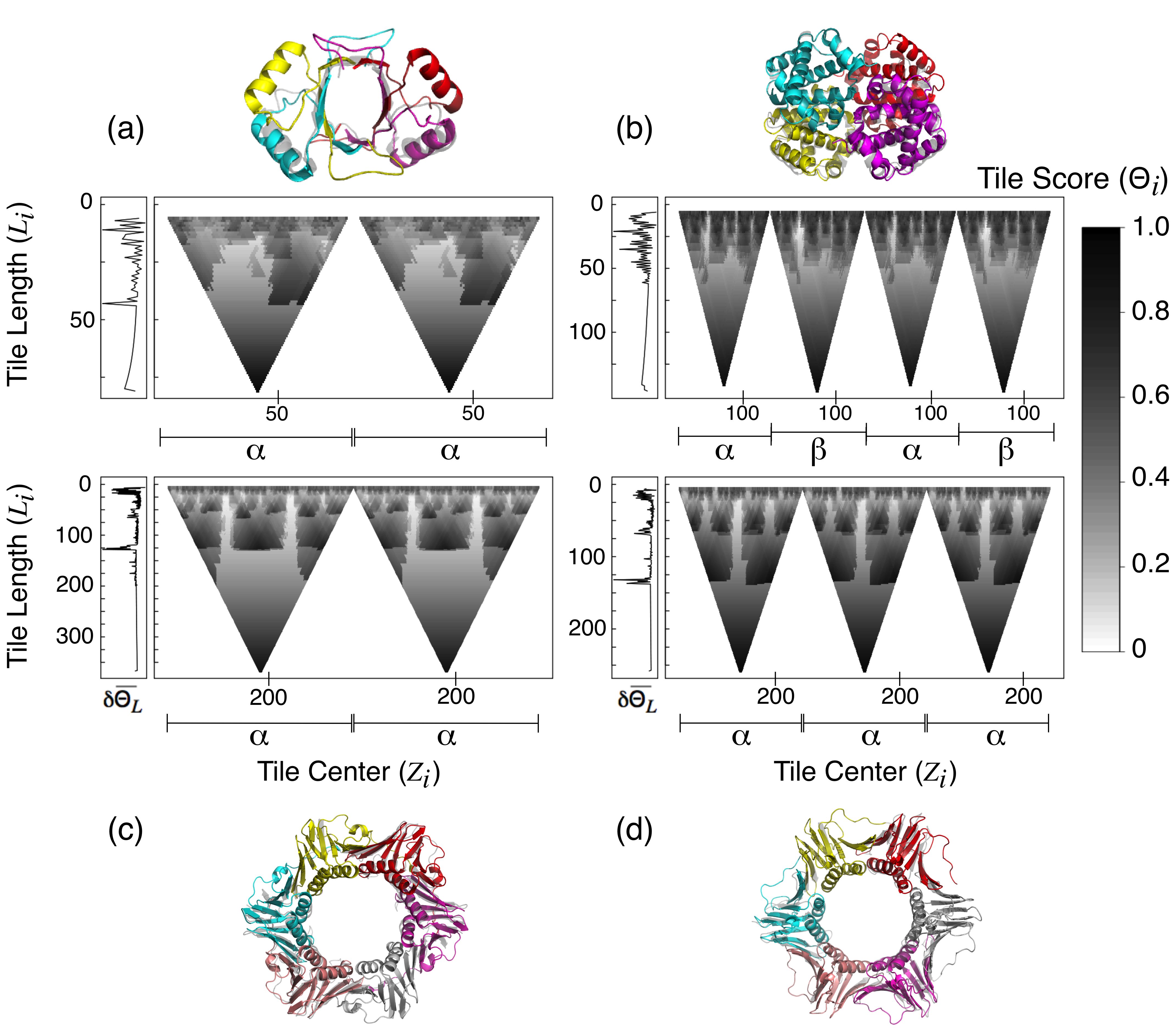}
	\caption{ Tiling quaternary complexes. The tiling profile is shown on grayscale, together with the $\delta\overline{\Theta_i}$ projected on the left. The structures of the native protein and example tilings are shown, using the same coloring scheme of Fig.2. The length ($L_{i}$) and center ($Z_{i}$) of the selected tile is: a) Homodimeric HPV-16 E2c (pdb:1r8p)  $L_i=43, Z_{i}=58.5$ b) Deoxy-Haemoglobin (pdb:2hhb)  $L_i=141, Z_{i}=71.5$ from chain A c) $\beta$-subunit of {\it Thermotoga maritima} DNA polymerase III (pdb:1vpk) $L_i=128, Z_{i}=297$ d) Processivity factor of {\it Saccharomyces cerevisiae} DNA polymerase-$\delta$ (pdb:1plq) $L_i=132, Z_{i}=190$}
	\label{fig:complexes}
\end{figure}
\medskip

Papillomavirus E2c-DNA binding protein is a remarkable model to study sequence specific recognition \cite{pmid20375284}. This domain is composed of two identical chains that come together forming in a $\beta$-barrel architecture that expose four $\alpha$-helices. The tiling procedure identifies a 81 residue fragment as the best scoring fragments, corresponding to the monomeric chains (Fig 6a). However, these can be further decomposed in tiles of $L_i=43$, covering about 90\% of the structural space. The best tile at this frequency corresponds to a $\beta\alpha\beta$ motif that intertwines in each monomer and together contribute half $\beta$-barrel (Fig 6a).

Haemoglobin (the helium atom of biology?) is the prime example of a symmetrical quaternary arrangement, a tetramer of $\alpha{_2}\beta_2$ chains. Figure 6b shows a regular tiling pattern in which four nearly identical regions can be distinguished. This highlights the long-established structural identity of the $\alpha$ and $\beta$ chains. As in the case of Myoglobin, no significant decomposition of the structure can be made with continuous fragments.

In occasions protein structures reveal geometrical chances and necessities of their history. Figure 6 show the structures of $\beta$-subunit of an archaeal DNA polymerase III (a homo-dimer, Fig 6c), together with the processivity factor of eukaryotic DNA polymerase-$\delta$ (a homo-trimer, Fig 6d). Tiling these quaternary complexes identifies the subunits and further point to similar characteristic frequencies of $L_i=128$ and $L_i=132$. In both cases the chosen tiles at their respective $L_{i}$ cover about 94\% of the structure of the complexes. It is apparent that a DNA clamp of this kind can be constructed with either two or three polypeptide chains, each containing three or two tiles, that pack in a sixfold rotational fashion \cite{pmid22483118}. This common tile can be further decomposed in 2 tiles of $L_i=65$ amino acids yet compromising about 10\% coverage. It is interesting to note that these smaller fragments get intertwined when forming a higher order structure, unlike any other of the maximal fragments identified.

\section*{Conclusions}

Foldable sequences with funneled landscapes are easier to find if the low energy structure is symmetric \cite{pmid8962034}. Modern natural philosophers appreciate the existence of symmetry as an emergent feature of the parsimony of nature, resulting from the limited modes of interaction between a small number of elementary parts assembling into higher order structures \cite{pmid10940245,pmidPeterFugue,pmid22615466,pmid12419661}. It is the inexact symmetries of biological molecules that are most striking \cite{pmid8962034,pmidPeterFugue}.  Subtle aperiodicities can give rise to big biological effects \cite{whatislife}, and thus their modulation can be at the core of the physiological workings of these ``frozen accidents''.

In order to detect and characterize repetitions in protein structures, we presented a simple scheme based on analyzing the distribution of suboptimal structural alignments of continuous fragments. The procedure identifies maximal fragments, those for which any extension occurs fewer times in the ensemble of solutions (Fig 1B). By counting the number of occurrences of non-overlapping fragments and having a good metric for the overall coverage, we defined a score that ranks how a structure can be tessellated with similar, though not identical, fragments. We found that in most cases there is a defined fragment length at which the coverage gained by the repetitions is highest, defining a characteristic frequency. In some cases there is a discrete collection of fragments that allows to unequivocally define a best phase. In these cases the repeat unit, the number of occurrences, and their boundaries can be confidently defined (Table S1). In other cases, there are several equivalent phases at the characteristic frequency, pointing to structures that can be considered almost periodic and where the definition of a basic tile must remain arbitrary (Fig 2, Table S1). This is a common theme in the cases of solenoidal proteins where different researchers have defined the repeat unit at distinct frequencies and phases \cite{pmid22923522}. Including other information beyond geometry could indicate if there is a {\it biologically} preferred phase, such as the characterization of insertion sites, the variability in orthologous sequences,  exon boundaries or folding mechanisms \cite{pmid23251375}.

	Proteins in which the repeats pack symmetrically against each other but do not translate along an axis can form closed structures. The fragmenting and tiling approach can be readily applied to such topologies like barrels, propellers, trefoils, and so on. Within these we can distinguish nested repeating units and even resolve fine geometrical differences (Fig. 4, Table S1).
If the fundamental tiles are arranged symmetrically, then there must be larger tiles which are multiples of these basic tiles. These higher order tiles appear as additional maxima of $\Theta_{L_i}$ towards larger $L_i$ as compared to the basic tile. This hierarchical nesting of tiles can be captured by a tessellation score that is computed in the following way. For each tile length $L_i$ take the maximum tile score $\Theta_i$ (e.g. the maximum score for a particular $L_i$ in Figure 1b) and take the average over all $L$. This  {\it tileability} score ($\Xi$) is 1.0 for the homogeneous model, approaches $1$ for highly regular structures like $\alpha$-helices and goes to zero for non-repetitive structures. In Figure 7 a variety of proteins are ranked by their respective tileability score $\Xi$ (Table S1). The largest value of $\Xi$ is obtained for a long $\alpha$-helix from a coiled coil. The helix is followed by several solenoidal proteins with the most regular designed proteins ranking higher than the more irregular natural ones. These structures are followed by repetitive proteins with an overall globular shape. At the end of this scale we find typical globular domains that do not have any periodicity larger than a few residues. We note that not all members of a particular topology group together, they rather get segregated according to the irregularities they display (Fig. 7). 

\medskip
	\begin{figure}
\centering
	\includegraphics[width=0.9\textwidth]{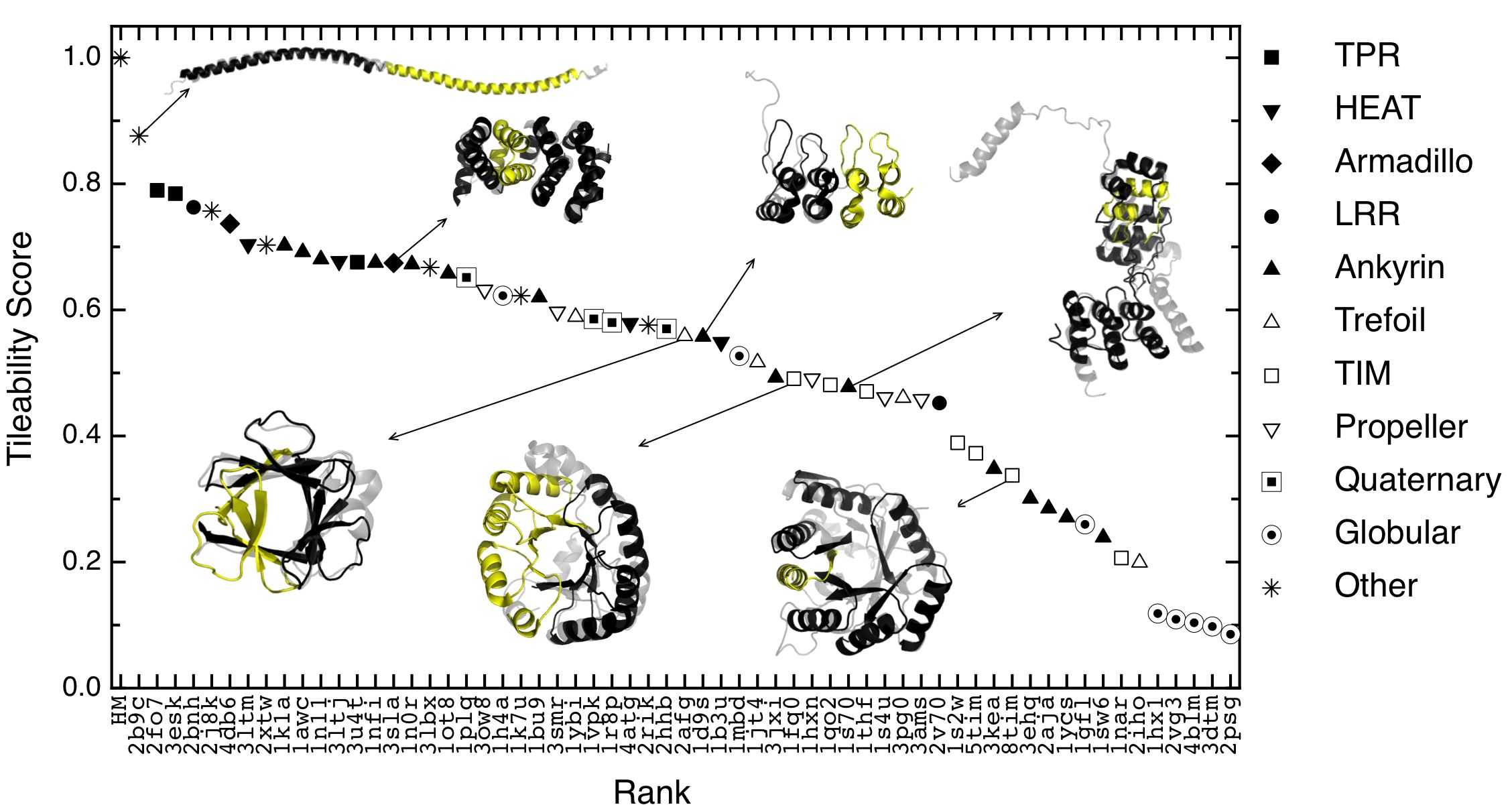}
	\caption{ Tileability of protein structures. The tiling procedure was applied to the protein models (indicated with their PDB code, HM: homogeneous model) and ranked according to their tileability score $\Xi$. Example tessellations are shown with the tile-unit colored yellow and the copies colored black, superimposed to the native structure in gray. Filled symbols: solenoidal repeat proteins. Empty symbols: globular repeat proteins.}
	\label{fig:complexes}
\end{figure}
\medskip

The same tiling procedure can be applied at the level of protein complexes, analyzing the details of how fragment copies between chains cover the structural space. At this level we found that the best tiles often correspond with the monomeric chains or classical globular domains within them. However interesting exceptions can mark chains that can be further decomposed into smaller units (Fig 6). It will be appealing to extend this now limited survey and characterize how frequently the distribution of geometric tiles coincide with the polypeptide chains, globular domains, exons boundaries, foldons or motifs.

	It is tempting to speculate about the functional consequences that the symmetrical distribution of similar fragments can have at different length scales. Energy Landscape Theory  {\it modus operandi} appreciates that packing subunits in symmetrically equivalent ways give rise to structures with similar free energies, allowing multiple funnels to coexist in the energy landscape \cite{pmid15701699} and small perturbations to switch between these states \cite{pmid19436496}. Symmetry has been pointed to be key in other functional phenomena such as folding cooperativity, multiple ligand binding, thermodynamic stability, coding compression, and finite assembly  \cite{pmid22615466,pmid10940245}. Symmetric organization is an easy (and perhaps unavoidable) way for allostery to emerge \cite{pmid14343300,pmid18075577}. Repetitions with point symmetries give rise to closed arrays such as barrels and the like at the tertiary level, and rings or polyhedra at the quaternary level. Helical symmetries form solenoids at the tertiary level that correspond with tubular organizations at the quaternary level. Nucleation and capping of these repeating arrays is often pointed to be critical to their physiological behavior both at the tertiary and quaternary levels. Potentially unbounded periodicity may require other mechanisms to terminate growth. It is thus not surprising that physiological workings and pathological states are the result of aggregation of similar fragments, such as cytoskelton dynamics \cite{pmid21876141}, epigenetic phenomena \cite{pmid19606595}, sickle-cell anemia \cite{pmid15395398} and amyloid-related processes \cite{pmid19411847}.

\par The organization of protein molecules can be appreciated at many levels, from amino acid sequence motifs to dynamic interacting networks of thousands of components \cite{pmid21876141}. As the relevant contributions of the physical forces change at different length and time scales, the organizational agencies at each level will necessarily differ, but some common principles may underlie. The concepts postulated by Energy Landscape Theory can be a guide in such search\cite{pmid1749933,pmid11875198,pmid20819242}

\section*{Dedication}
	Natural protein molecules are indeed very peculiar polymers. The works of Peter G. Wolynes, to whom we celebrate his third 20th birthday in this volume, transcended fields and provided us with deep impressions and equations to appreciate nature's beauty and comprehend its rich complexity. Enumerating his vast production and scientific achievements does not come close to the illuminating experience the lucky of us had in investigating with him. To Peter then we raise our Martini in {\it funneled glasses} and repeat `` !`Salud! ''.

\acknowledgement

This work was supported by the Consejo Nacional de Investigaciones Cient\'ificas y T\'ecnicas de Argentina (CONICET), the Agencia Nacional de Promoci\'on Cient\'ifica y Tecnol\'ogica (ANPCyT), and by FWF Austria, grant number P21294-B12. RGP and RE hold fellowships from CONICET, IES and DUF are Career Investigators.

\subsection*{TOC image}

\medskip
	\begin{figure}
\centering
	\includegraphics[width=0.3\textwidth]{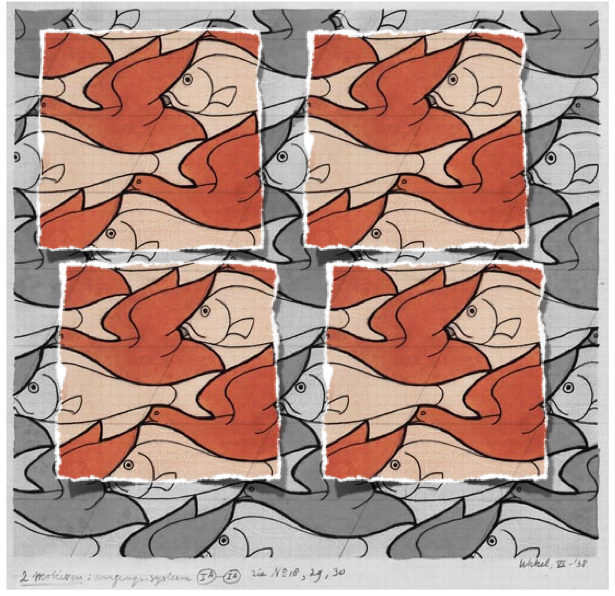}
	\label{fig:toc}
\end{figure}
\medskip

%

\begin{suppinfo}
Supporting information includes Table S1 with the tile and tessellation parameters for all the surveyed structures, the derivation of the homogeneous model and figures of tilings and tessellations of example proteins.

\end{suppinfo}

\bibliography{Bibliography.bib}

\providecommand*\mcitethebibliography{\thebibliography}
\csname @ifundefined\endcsname{endmcitethebibliography}
  {\let\endmcitethebibliography\endthebibliography}{}
\begin{mcitethebibliography}{70}
\providecommand*\natexlab[1]{#1}
\providecommand*\mciteSetBstSublistMode[1]{}
\providecommand*\mciteSetBstMaxWidthForm[2]{}
\providecommand*\mciteBstWouldAddEndPuncttrue
  {\def\EndOfBibitem{\unskip.}}
\providecommand*\mciteBstWouldAddEndPunctfalse
  {\let\EndOfBibitem\relax}
\providecommand*\mciteSetBstMidEndSepPunct[3]{}
\providecommand*\mciteSetBstSublistLabelBeginEnd[3]{}
\providecommand*\EndOfBibitem{}
\mciteSetBstSublistMode{f}
\mciteSetBstMaxWidthForm{subitem}{(\alph{mcitesubitemcount})}
\mciteSetBstSublistLabelBeginEnd
  {\mcitemaxwidthsubitemform\space}
  {\relax}
  {\relax}

\bibitem[Bryngelson et~al.(1995)Bryngelson, Onuchic, Socci, and
  Wolynes]{pmid7784423}
Bryngelson,~J.~D.; Onuchic,~J.~N.; Socci,~N.~D.; Wolynes,~P.~G. {{F}unnels,
  Pathways, and the Energy Landscape of Protein Folding: a Synthesis}.
  \emph{Proteins} \textbf{1995}, \emph{21}, 167--195\relax
\mciteBstWouldAddEndPuncttrue
\mciteSetBstMidEndSepPunct{\mcitedefaultmidpunct}
{\mcitedefaultendpunct}{\mcitedefaultseppunct}\relax
\EndOfBibitem
\bibitem[Wolynes(2005)]{pmid16934172}
Wolynes,~P.~G. Recent Successes of the Energy Landscape Theory of Protein
  Folding and Function. \emph{Q Rev Biophys} \textbf{2005}, \emph{38},
  405--10\relax
\mciteBstWouldAddEndPuncttrue
\mciteSetBstMidEndSepPunct{\mcitedefaultmidpunct}
{\mcitedefaultendpunct}{\mcitedefaultseppunct}\relax
\EndOfBibitem
\bibitem[Wolynes(2005)]{pmid15664893}
Wolynes,~P.~G. {{E}nergy Landscapes and Solved Protein-folding Problems}.
  \emph{Philos Transact A Math Phys Eng Sci} \textbf{2005}, \emph{363},
  453--464\relax
\mciteBstWouldAddEndPuncttrue
\mciteSetBstMidEndSepPunct{\mcitedefaultmidpunct}
{\mcitedefaultendpunct}{\mcitedefaultseppunct}\relax
\EndOfBibitem
\bibitem[Davtyan et~al.(2012)Davtyan, Schafer, Zheng, Clementi, Wolynes, and
  Papoian]{pmid22545654}
Davtyan,~A.; Schafer,~N.~P.; Zheng,~W.; Clementi,~C.; Wolynes,~P.~G.;
  Papoian,~G.~A. {{A}{W}{S}{E}{M}-{M}{D}: Protein Structure Prediction Using
  Coarse-grained Physical Potentials and Bioinformatically Based Local
  Structure Biasing}. \emph{J Phys Chem B} \textbf{2012}, \emph{116},
  8494--8503\relax
\mciteBstWouldAddEndPuncttrue
\mciteSetBstMidEndSepPunct{\mcitedefaultmidpunct}
{\mcitedefaultendpunct}{\mcitedefaultseppunct}\relax
\EndOfBibitem
\bibitem[Zheng et~al.(2012)Zheng, Schafer, Davtyan, Papoian, and
  Wolynes]{pmid23129648}
Zheng,~W.; Schafer,~N.~P.; Davtyan,~A.; Papoian,~G.~A.; Wolynes,~P.~G.
  {{P}redictive Energy Landscapes for Protein-protein Association}. \emph{Proc.
  Natl. Acad. Sci. U.S.A.} \textbf{2012}, \emph{109}, 19244--19249\relax
\mciteBstWouldAddEndPuncttrue
\mciteSetBstMidEndSepPunct{\mcitedefaultmidpunct}
{\mcitedefaultendpunct}{\mcitedefaultseppunct}\relax
\EndOfBibitem
\bibitem[Wolynes et~al.(2012)Wolynes, Eaton, and Fersht]{pmid23112193}
Wolynes,~P.~G.; Eaton,~W.~A.; Fersht,~A.~R. {{C}hemical Physics of Protein
  Folding}. \emph{Proc. Natl. Acad. Sci. U.S.A.} \textbf{2012}, \emph{109},
  17770--17771\relax
\mciteBstWouldAddEndPuncttrue
\mciteSetBstMidEndSepPunct{\mcitedefaultmidpunct}
{\mcitedefaultendpunct}{\mcitedefaultseppunct}\relax
\EndOfBibitem
\bibitem[Oliveberg and Wolynes(2005)Oliveberg, and Wolynes]{pmid16780604}
Oliveberg,~M.; Wolynes,~P.~G. {{T}he Experimental Survey of Protein-folding
  Energy Landscapes}. \emph{Q. Rev. Biophys.} \textbf{2005}, \emph{38},
  245--288\relax
\mciteBstWouldAddEndPuncttrue
\mciteSetBstMidEndSepPunct{\mcitedefaultmidpunct}
{\mcitedefaultendpunct}{\mcitedefaultseppunct}\relax
\EndOfBibitem
\bibitem[Bryngelson and Wolynes(1987)Bryngelson, and Wolynes]{pmid3478708}
Bryngelson,~J.~D.; Wolynes,~P.~G. {{S}pin Glasses and the Statistical Mechanics
  of Protein Folding}. \emph{Proc. Natl. Acad. Sci. U.S.A.} \textbf{1987},
  \emph{84}, 7524--7528\relax
\mciteBstWouldAddEndPuncttrue
\mciteSetBstMidEndSepPunct{\mcitedefaultmidpunct}
{\mcitedefaultendpunct}{\mcitedefaultseppunct}\relax
\EndOfBibitem
\bibitem[Weiss et~al.(2000)Weiss, Jimenez-Montano, and Herzel]{pmid10988023}
Weiss,~O.; Jimenez-Montano,~M.~A.; Herzel,~H. {{I}nformation Content of Protein
  Sequences}. \emph{J. Theor. Biol.} \textbf{2000}, \emph{206}, 379--386\relax
\mciteBstWouldAddEndPuncttrue
\mciteSetBstMidEndSepPunct{\mcitedefaultmidpunct}
{\mcitedefaultendpunct}{\mcitedefaultseppunct}\relax
\EndOfBibitem
\bibitem[Wolynes(1996)]{pmid8962034}
Wolynes,~P.~G. {{S}ymmetry and the Energy Landscapes of Biomolecules}.
  \emph{Proc. Natl. Acad. Sci. U.S.A.} \textbf{1996}, \emph{93},
  14249--14255\relax
\mciteBstWouldAddEndPuncttrue
\mciteSetBstMidEndSepPunct{\mcitedefaultmidpunct}
{\mcitedefaultendpunct}{\mcitedefaultseppunct}\relax
\EndOfBibitem
\bibitem[Panchenko et~al.(1996)Panchenko, Luthey-Schulten, and
  Wolynes]{pmid8700876}
Panchenko,~A.~R.; Luthey-Schulten,~Z.; Wolynes,~P.~G. {{F}oldons, Protein
  Structural Modules, and Exons}. \emph{Proc. Natl. Acad. Sci. U.S.A.}
  \textbf{1996}, \emph{93}, 2008--2013\relax
\mciteBstWouldAddEndPuncttrue
\mciteSetBstMidEndSepPunct{\mcitedefaultmidpunct}
{\mcitedefaultendpunct}{\mcitedefaultseppunct}\relax
\EndOfBibitem
\bibitem[Wales(1998)]{pmidWalessym98}
Wales,~D.~J. Symmetry, Near-symmetry and Energetics. \emph{Chem. Phys. Lett.}
  \textbf{1998}, \emph{285}, 330--336\relax
\mciteBstWouldAddEndPuncttrue
\mciteSetBstMidEndSepPunct{\mcitedefaultmidpunct}
{\mcitedefaultendpunct}{\mcitedefaultseppunct}\relax
\EndOfBibitem
\bibitem[Ferreiro and Wolynes(2008)Ferreiro, and Wolynes]{pmid18632565}
Ferreiro,~D.~U.; Wolynes,~P.~G. {{T}he Capillarity Picture and the Kinetics of
  One-dimensional Protein Folding}. \emph{Proc. Natl. Acad. Sci. U.S.A.}
  \textbf{2008}, \emph{105}, 9853--9854\relax
\mciteBstWouldAddEndPuncttrue
\mciteSetBstMidEndSepPunct{\mcitedefaultmidpunct}
{\mcitedefaultendpunct}{\mcitedefaultseppunct}\relax
\EndOfBibitem
\bibitem[Itoh and Sasai(2009)Itoh, and Sasai]{pmid19368477}
Itoh,~K.; Sasai,~M. {{M}ultidimensional Theory of Protein Folding}. \emph{J
  Chem Phys} \textbf{2009}, \emph{130}, 145104\relax
\mciteBstWouldAddEndPuncttrue
\mciteSetBstMidEndSepPunct{\mcitedefaultmidpunct}
{\mcitedefaultendpunct}{\mcitedefaultseppunct}\relax
\EndOfBibitem
\bibitem[Luo and Nijveen(2013)Luo, and Nijveen]{pmid23418055}
Luo,~H.; Nijveen,~H. Understanding and Identifying Amino Acid Repeats.
  \emph{Brief Bioinform In press. doi: 10.1093/bib/bbt003} \textbf{2013},
  \relax
\mciteBstWouldAddEndPunctfalse
\mciteSetBstMidEndSepPunct{\mcitedefaultmidpunct}
{}{\mcitedefaultseppunct}\relax
\EndOfBibitem
\bibitem[Kajava(2012)]{pmid21884799}
Kajava,~A.~V. Tandem Repeats in Proteins: from Sequence to Structure. \emph{J
  Struct Biol} \textbf{2012}, \emph{179}, 279--88\relax
\mciteBstWouldAddEndPuncttrue
\mciteSetBstMidEndSepPunct{\mcitedefaultmidpunct}
{\mcitedefaultendpunct}{\mcitedefaultseppunct}\relax
\EndOfBibitem
\bibitem[Shih and Hwang(2004)Shih, and Hwang]{pmid15229884}
Shih,~E.~S.; Hwang,~M.~J. {{A}lternative Alignments from Comparison of Protein
  Structures}. \emph{Proteins} \textbf{2004}, \emph{56}, 519--527\relax
\mciteBstWouldAddEndPuncttrue
\mciteSetBstMidEndSepPunct{\mcitedefaultmidpunct}
{\mcitedefaultendpunct}{\mcitedefaultseppunct}\relax
\EndOfBibitem
\bibitem[Abraham et~al.(2008)Abraham, Rocha, and Pothier]{pmid18487242}
Abraham,~A.~L.; Rocha,~E.~P.; Pothier,~J. {{S}welfe: a Detector of Internal
  Repeats in Sequences and Structures}. \emph{Bioinformatics} \textbf{2008},
  \emph{24}, 1536--1537\relax
\mciteBstWouldAddEndPuncttrue
\mciteSetBstMidEndSepPunct{\mcitedefaultmidpunct}
{\mcitedefaultendpunct}{\mcitedefaultseppunct}\relax
\EndOfBibitem
\bibitem[Murray et~al.(2004)Murray, Taylor, and Thornton]{pmid15340924}
Murray,~K.~B.; Taylor,~W.~R.; Thornton,~J.~M. {{T}oward the Detection and
  Validation of Repeats in Protein Structure}. \emph{Proteins} \textbf{2004},
  \emph{57}, 365--380\relax
\mciteBstWouldAddEndPuncttrue
\mciteSetBstMidEndSepPunct{\mcitedefaultmidpunct}
{\mcitedefaultendpunct}{\mcitedefaultseppunct}\relax
\EndOfBibitem
\bibitem[Taylor et~al.(2002)Taylor, Heringa, Baud, and Flores]{pmid11917144}
Taylor,~W.~R.; Heringa,~J.; Baud,~F.; Flores,~T.~P. A Fourier Analysis of
  Symmetry in Protein Structure. \emph{Protein Eng} \textbf{2002}, \emph{15},
  79--89\relax
\mciteBstWouldAddEndPuncttrue
\mciteSetBstMidEndSepPunct{\mcitedefaultmidpunct}
{\mcitedefaultendpunct}{\mcitedefaultseppunct}\relax
\EndOfBibitem
\bibitem[Walsh et~al.(2012)Walsh, Sirocco, Minervini, Di~Domenico, Ferrari, and
  Tosatto]{pmid22962341}
Walsh,~I.; Sirocco,~F.~G.; Minervini,~G.; Di~Domenico,~T.; Ferrari,~C.;
  Tosatto,~S.~C. {{R}{A}{P}{H}{A}{E}{L}: Recognition, Periodicity and Insertion
  Assignment of Solenoid Protein Structures}. \emph{Bioinformatics}
  \textbf{2012}, \emph{28}, 3257--3264\relax
\mciteBstWouldAddEndPuncttrue
\mciteSetBstMidEndSepPunct{\mcitedefaultmidpunct}
{\mcitedefaultendpunct}{\mcitedefaultseppunct}\relax
\EndOfBibitem
\bibitem[Marcotte et~al.(1999)Marcotte, Pellegrini, Yeates, and
  Eisenberg]{pmid10512723}
Marcotte,~E.~M.; Pellegrini,~M.; Yeates,~T.~O.; Eisenberg,~D. {{A} Census of
  Protein Repeats}. \emph{J. Mol. Biol.} \textbf{1999}, \emph{293},
  151--160\relax
\mciteBstWouldAddEndPuncttrue
\mciteSetBstMidEndSepPunct{\mcitedefaultmidpunct}
{\mcitedefaultendpunct}{\mcitedefaultseppunct}\relax
\EndOfBibitem
\bibitem[Schaper et~al.(2012)Schaper, Kajava, Hauser, and
  Anisimova]{pmid22923522}
Schaper,~E.; Kajava,~A.~V.; Hauser,~A.; Anisimova,~M. {{R}epeat or not
  Repeat?--{S}tatistical Validation of Tandem Repeat Prediction in Genomic
  Sequences}. \emph{Nucleic Acids Res.} \textbf{2012}, \emph{40},
  10005--10017\relax
\mciteBstWouldAddEndPuncttrue
\mciteSetBstMidEndSepPunct{\mcitedefaultmidpunct}
{\mcitedefaultendpunct}{\mcitedefaultseppunct}\relax
\EndOfBibitem
\bibitem[Sippl and Wiederstein(2012)Sippl, and Wiederstein]{pmid22483118}
Sippl,~M.~J.; Wiederstein,~M. {{D}etection of Spatial Correlations in Protein
  Structures and Molecular Complexes}. \emph{Structure} \textbf{2012},
  \emph{20}, 718--728\relax
\mciteBstWouldAddEndPuncttrue
\mciteSetBstMidEndSepPunct{\mcitedefaultmidpunct}
{\mcitedefaultendpunct}{\mcitedefaultseppunct}\relax
\EndOfBibitem
\bibitem[Sippl and Wiederstein(2008)Sippl, and Wiederstein]{pmid18174182}
Sippl,~M.~J.; Wiederstein,~M. {{A} Note on Difficult Structure Alignment
  Problems}. \emph{Bioinformatics} \textbf{2008}, \emph{24}, 426--427\relax
\mciteBstWouldAddEndPuncttrue
\mciteSetBstMidEndSepPunct{\mcitedefaultmidpunct}
{\mcitedefaultendpunct}{\mcitedefaultseppunct}\relax
\EndOfBibitem
\bibitem[Sippl(2008)]{pmid18227113}
Sippl,~M.~J. On Distance and Similarity in Fold Space. \emph{Bioinformatics}
  \textbf{2008}, \emph{24}, 872--3\relax
\mciteBstWouldAddEndPuncttrue
\mciteSetBstMidEndSepPunct{\mcitedefaultmidpunct}
{\mcitedefaultendpunct}{\mcitedefaultseppunct}\relax
\EndOfBibitem
\bibitem[Mosavi et~al.(2002)Mosavi, Minor, and Peng]{pmid12461176}
Mosavi,~L.~K.; Minor,~D.~L.; Peng,~Z.~Y. {{C}onsensus-derived Structural
  Determinants of the Ankyrin Repeat Motif}. \emph{Proc. Natl. Acad. Sci.
  U.S.A.} \textbf{2002}, \emph{99}, 16029--16034\relax
\mciteBstWouldAddEndPuncttrue
\mciteSetBstMidEndSepPunct{\mcitedefaultmidpunct}
{\mcitedefaultendpunct}{\mcitedefaultseppunct}\relax
\EndOfBibitem
\bibitem[Wolynes(1997)]{pmid9177189}
Wolynes,~P.~G. Folding Funnels and Energy Landscapes of Larger Proteins Within
  the Capillarity Approximation. \emph{Proc Natl Acad Sci U S A} \textbf{1997},
  \emph{94}, 6170--5\relax
\mciteBstWouldAddEndPuncttrue
\mciteSetBstMidEndSepPunct{\mcitedefaultmidpunct}
{\mcitedefaultendpunct}{\mcitedefaultseppunct}\relax
\EndOfBibitem
\bibitem[Ferreiro et~al.(2008)Ferreiro, Walczak, Komives, and
  Wolynes]{pmid18483553}
Ferreiro,~D.~U.; Walczak,~A.~M.; Komives,~E.~A.; Wolynes,~P.~G. {{T}he Energy
  Landscapes of Repeat-containing Proteins: Topology, Cooperativity, and the
  Folding Funnels of One-dimensional Architectures}. \emph{PLoS Comput. Biol.}
  \textbf{2008}, \emph{4}, e1000070\relax
\mciteBstWouldAddEndPuncttrue
\mciteSetBstMidEndSepPunct{\mcitedefaultmidpunct}
{\mcitedefaultendpunct}{\mcitedefaultseppunct}\relax
\EndOfBibitem
\bibitem[Ferreiro and Komives(2010)Ferreiro, and Komives]{pmid20055496}
Ferreiro,~D.~U.; Komives,~E.~A. Molecular Mechanisms of System Control of
  NF-kappaB Signaling by IkappaBalpha. \emph{Biochemistry} \textbf{2010},
  \emph{49}, 1560--7\relax
\mciteBstWouldAddEndPuncttrue
\mciteSetBstMidEndSepPunct{\mcitedefaultmidpunct}
{\mcitedefaultendpunct}{\mcitedefaultseppunct}\relax
\EndOfBibitem
\bibitem[DeVries et~al.(2011)DeVries, Ferreiro, Sanchez, and
  Komives]{pmid21329696}
DeVries,~I.; Ferreiro,~D.~U.; Sanchez,~I.~E.; Komives,~E.~A. {{F}olding
  Kinetics of the Cooperatively Folded Subdomain of the {I}kappa{B}alpha
  Ankyrin Repeat Domain}. \emph{J. Mol. Biol.} \textbf{2011}, \emph{408},
  163--176\relax
\mciteBstWouldAddEndPuncttrue
\mciteSetBstMidEndSepPunct{\mcitedefaultmidpunct}
{\mcitedefaultendpunct}{\mcitedefaultseppunct}\relax
\EndOfBibitem
\bibitem[Ferreiro et~al.(2007)Ferreiro, Cervantes, Truhlar, Cho, Wolynes, and
  Komives]{pmid17174335}
Ferreiro,~D.~U.; Cervantes,~C.~F.; Truhlar,~S.~M.; Cho,~S.~S.; Wolynes,~P.~G.;
  Komives,~E.~A. {{S}tabilizing {I}kappa{B}alpha by "consensus" Design}.
  \emph{J. Mol. Biol.} \textbf{2007}, \emph{365}, 1201--1216\relax
\mciteBstWouldAddEndPuncttrue
\mciteSetBstMidEndSepPunct{\mcitedefaultmidpunct}
{\mcitedefaultendpunct}{\mcitedefaultseppunct}\relax
\EndOfBibitem
\bibitem[Muraki et~al.(2002)Muraki, Ishimura, and Harata]{pmid11853952}
Muraki,~M.; Ishimura,~M.; Harata,~K. {{I}nteractions of Wheat-germ Agglutinin
  with {G}lc{N}{A}c Beta 1,6{G}al Sequence}. \emph{Biochim. Biophys. Acta}
  \textbf{2002}, \emph{1569}, 10--20\relax
\mciteBstWouldAddEndPuncttrue
\mciteSetBstMidEndSepPunct{\mcitedefaultmidpunct}
{\mcitedefaultendpunct}{\mcitedefaultseppunct}\relax
\EndOfBibitem
\bibitem[Haigis et~al.(2002)Haigis, Haag, and Raines]{pmid12032252}
Haigis,~M.~C.; Haag,~E.~S.; Raines,~R.~T. {{E}volution of Ribonuclease
  Inhibitor by Exon Duplication}. \emph{Mol. Biol. Evol.} \textbf{2002},
  \emph{19}, 959--963\relax
\mciteBstWouldAddEndPuncttrue
\mciteSetBstMidEndSepPunct{\mcitedefaultmidpunct}
{\mcitedefaultendpunct}{\mcitedefaultseppunct}\relax
\EndOfBibitem
\bibitem[Groves et~al.(1999)Groves, Hanlon, Turowski, Hemmings, and
  Barford]{pmid9989501}
Groves,~M.~R.; Hanlon,~N.; Turowski,~P.; Hemmings,~B.~A.; Barford,~D. The
  Structure of the Protein Phosphatase 2A PR65/A Subunit Reveals the
  Conformation of Its 15 Tandemly Repeated HEAT Motifs. \emph{Cell}
  \textbf{1999}, \emph{96}, 99--110\relax
\mciteBstWouldAddEndPuncttrue
\mciteSetBstMidEndSepPunct{\mcitedefaultmidpunct}
{\mcitedefaultendpunct}{\mcitedefaultseppunct}\relax
\EndOfBibitem
\bibitem[Nagano et~al.(2002)Nagano, Orengo, and Thornton]{pmid12206759}
Nagano,~N.; Orengo,~C.~A.; Thornton,~J.~M. One Fold with Many Functions: the
  Evolutionary Relationships Between TIM Barrel Families Based on their
  Sequences, Structures and Functions. \emph{J Mol Biol} \textbf{2002},
  \emph{321}, 741--65\relax
\mciteBstWouldAddEndPuncttrue
\mciteSetBstMidEndSepPunct{\mcitedefaultmidpunct}
{\mcitedefaultendpunct}{\mcitedefaultseppunct}\relax
\EndOfBibitem
\bibitem[Soding et~al.(2006)Soding, Remmert, and Biegert]{pmid16844977}
Soding,~J.; Remmert,~M.; Biegert,~A. {{H}{H}rep: De Novo Protein Repeat
  Detection and the Origin of {T}{I}{M} Barrels}. \emph{Nucleic Acids Res.}
  \textbf{2006}, \emph{34}, W137--142\relax
\mciteBstWouldAddEndPuncttrue
\mciteSetBstMidEndSepPunct{\mcitedefaultmidpunct}
{\mcitedefaultendpunct}{\mcitedefaultseppunct}\relax
\EndOfBibitem
\bibitem[Fulop and Jones(1999)Fulop, and Jones]{pmid10607670}
Fulop,~V.; Jones,~D.~T. {{B}eta Propellers: Structural Rigidity and Functional
  Diversity}. \emph{Curr. Opin. Struct. Biol.} \textbf{1999}, \emph{9},
  715--721\relax
\mciteBstWouldAddEndPuncttrue
\mciteSetBstMidEndSepPunct{\mcitedefaultmidpunct}
{\mcitedefaultendpunct}{\mcitedefaultseppunct}\relax
\EndOfBibitem
\bibitem[Neer et~al.(1994)Neer, Schmidt, Nambudripad, and Smith]{pmid8090199}
Neer,~E.~J.; Schmidt,~C.~J.; Nambudripad,~R.; Smith,~T.~F. The Ancient
  Regulatory-protein Family of WD-repeat Proteins. \emph{Nature} \textbf{1994},
  \emph{371}, 297--300\relax
\mciteBstWouldAddEndPuncttrue
\mciteSetBstMidEndSepPunct{\mcitedefaultmidpunct}
{\mcitedefaultendpunct}{\mcitedefaultseppunct}\relax
\EndOfBibitem
\bibitem[Pauling et~al.(1951)Pauling, Corey, and Branson]{pmid14816373}
Pauling,~L.; Corey,~R.~B.; Branson,~H.~R. The Structure of Proteins; Two
  Hydrogen-bonded Helical Configurations of the Polypeptide Chain. \emph{Proc
  Natl Acad Sci U S A} \textbf{1951}, \emph{37}, 205--11\relax
\mciteBstWouldAddEndPuncttrue
\mciteSetBstMidEndSepPunct{\mcitedefaultmidpunct}
{\mcitedefaultendpunct}{\mcitedefaultseppunct}\relax
\EndOfBibitem
\bibitem[Pauling and Corey(1951)Pauling, and Corey]{pmid14834147}
Pauling,~L.; Corey,~R.~B. The Pleated Sheet, a New Layer Configuration of
  Polypeptide Chains. \emph{Proc Natl Acad Sci U S A} \textbf{1951}, \emph{37},
  251--6\relax
\mciteBstWouldAddEndPuncttrue
\mciteSetBstMidEndSepPunct{\mcitedefaultmidpunct}
{\mcitedefaultendpunct}{\mcitedefaultseppunct}\relax
\EndOfBibitem
\bibitem[Moult et~al.(2011)Moult, Fidelis, Kryshtafovych, and
  Tramontano]{pmid21997831}
Moult,~J.; Fidelis,~K.; Kryshtafovych,~A.; Tramontano,~A. Critical Assessment
  of Methods of Protein Structure Prediction (CASP)--round IX. \emph{Proteins}
  \textbf{2011}, \emph{79 Suppl 10}, 1--5\relax
\mciteBstWouldAddEndPuncttrue
\mciteSetBstMidEndSepPunct{\mcitedefaultmidpunct}
{\mcitedefaultendpunct}{\mcitedefaultseppunct}\relax
\EndOfBibitem
\bibitem[Hegler et~al.(2009)Hegler, L{\"a}tzer, Shehu, Clementi, and
  Wolynes]{pmid19706384}
Hegler,~J.~A.; L{\"a}tzer,~J.; Shehu,~A.; Clementi,~C.; Wolynes,~P.~G.
  Restriction Versus Guidance in Protein Structure Prediction. \emph{Proc Natl
  Acad Sci U S A} \textbf{2009}, \emph{106}, 15302--7\relax
\mciteBstWouldAddEndPuncttrue
\mciteSetBstMidEndSepPunct{\mcitedefaultmidpunct}
{\mcitedefaultendpunct}{\mcitedefaultseppunct}\relax
\EndOfBibitem
\bibitem[Simons et~al.(1997)Simons, Kooperberg, Huang, and Baker]{pmid9149153}
Simons,~K.~T.; Kooperberg,~C.; Huang,~E.; Baker,~D. Assembly of Protein
  Tertiary Structures from Fragments with Similar Local Sequences Using
  Simulated Annealing and Bayesian Scoring Functions. \emph{J Mol Biol}
  \textbf{1997}, \emph{268}, 209--25\relax
\mciteBstWouldAddEndPuncttrue
\mciteSetBstMidEndSepPunct{\mcitedefaultmidpunct}
{\mcitedefaultendpunct}{\mcitedefaultseppunct}\relax
\EndOfBibitem
\bibitem[Truscott(2011)]{pmid21272671}
Truscott,~R. J.~W. Macromolecular Deterioration as the Ultimate Constraint on
  Human Lifespan. \emph{Ageing Res Rev} \textbf{2011}, \emph{10},
  397--403\relax
\mciteBstWouldAddEndPuncttrue
\mciteSetBstMidEndSepPunct{\mcitedefaultmidpunct}
{\mcitedefaultendpunct}{\mcitedefaultseppunct}\relax
\EndOfBibitem
\bibitem[Frauenfelder et~al.(2003)Frauenfelder, McMahon, and
  Fenimore]{pmid12861080}
Frauenfelder,~H.; McMahon,~B.~H.; Fenimore,~P.~W. Myoglobin: the Hydrogen Atom
  of Biology and a Paradigm of Complexity. \emph{Proc Natl Acad Sci U S A}
  \textbf{2003}, \emph{100}, 8615--7\relax
\mciteBstWouldAddEndPuncttrue
\mciteSetBstMidEndSepPunct{\mcitedefaultmidpunct}
{\mcitedefaultendpunct}{\mcitedefaultseppunct}\relax
\EndOfBibitem
\bibitem[Kendrew et~al.(1958)Kendrew, Bodo, Dintzis, Parrish, Wyckoff, and
  Phillips]{pmid13517261}
Kendrew,~J.; Bodo,~G.; Dintzis,~H.~M.; Parrish,~R.~G.; Wyckoff,~H.;
  Phillips,~D.~C. A Three-dimensional Model of the Myoglobin Molecule Obtained
  by X-ray Analysis. \emph{Nature} \textbf{1958}, \emph{181}, 662--6\relax
\mciteBstWouldAddEndPuncttrue
\mciteSetBstMidEndSepPunct{\mcitedefaultmidpunct}
{\mcitedefaultendpunct}{\mcitedefaultseppunct}\relax
\EndOfBibitem
\bibitem[Orm{\"o} et~al.(1996)Orm{\"o}, Cubitt, Kallio, Gross, Tsien, and
  Remington]{pmid8703075}
Orm{\"o},~M.; Cubitt,~A.~B.; Kallio,~K.; Gross,~L.~A.; Tsien,~R.~Y.;
  Remington,~S.~J. Crystal Structure of the Aequorea Victoria Green Fluorescent
  Protein. \emph{Science} \textbf{1996}, \emph{273}, 1392--5\relax
\mciteBstWouldAddEndPuncttrue
\mciteSetBstMidEndSepPunct{\mcitedefaultmidpunct}
{\mcitedefaultendpunct}{\mcitedefaultseppunct}\relax
\EndOfBibitem
\bibitem[Santos et~al.(2004)Santos, Gebhard, Risso, Ferreyra, Rossi, and
  Erm{\'a}cora]{pmid14769049}
Santos,~J.; Gebhard,~L.~G.; Risso,~V.~A.; Ferreyra,~R.~G.; Rossi,~J. P. F.~C.;
  Erm{\'a}cora,~M.~R. Folding of an Abridged Beta-lactamase.
  \emph{Biochemistry} \textbf{2004}, \emph{43}, 1715--23\relax
\mciteBstWouldAddEndPuncttrue
\mciteSetBstMidEndSepPunct{\mcitedefaultmidpunct}
{\mcitedefaultendpunct}{\mcitedefaultseppunct}\relax
\EndOfBibitem
\bibitem[Goodsell and Olson(2000)Goodsell, and Olson]{pmid10940245}
Goodsell,~D.~S.; Olson,~A.~J. {{S}tructural Symmetry and Protein Function}.
  \emph{Annu Rev Biophys Biomol Struct} \textbf{2000}, \emph{29},
  105--153\relax
\mciteBstWouldAddEndPuncttrue
\mciteSetBstMidEndSepPunct{\mcitedefaultmidpunct}
{\mcitedefaultendpunct}{\mcitedefaultseppunct}\relax
\EndOfBibitem
\bibitem[Swapna et~al.(2012)Swapna, Srikeerthana, and Srinivasan]{pmid22629324}
Swapna,~L.~S.; Srikeerthana,~K.; Srinivasan,~N. {{E}xtent of Structural
  Asymmetry in Homodimeric Proteins: Prevalence and Relevance}. \emph{PLoS ONE}
  \textbf{2012}, \emph{7}, e36688\relax
\mciteBstWouldAddEndPuncttrue
\mciteSetBstMidEndSepPunct{\mcitedefaultmidpunct}
{\mcitedefaultendpunct}{\mcitedefaultseppunct}\relax
\EndOfBibitem
\bibitem[Hashimoto and Panchenko(2010)Hashimoto, and Panchenko]{pmid21048085}
Hashimoto,~K.; Panchenko,~A.~R. {{M}echanisms of Protein Oligomerization, the
  Critical Role of Insertions and Deletions in Maintaining Different Oligomeric
  States}. \emph{Proc. Natl. Acad. Sci. U.S.A.} \textbf{2010}, \emph{107},
  20352--20357\relax
\mciteBstWouldAddEndPuncttrue
\mciteSetBstMidEndSepPunct{\mcitedefaultmidpunct}
{\mcitedefaultendpunct}{\mcitedefaultseppunct}\relax
\EndOfBibitem
\bibitem[S{\'a}nchez et~al.(2010)S{\'a}nchez, Ferreiro, Dellarole, and
  de~Prat-Gay]{pmid20375284}
S{\'a}nchez,~I.~E.; Ferreiro,~D.~U.; Dellarole,~M.; de~Prat-Gay,~G.
  Experimental Snapshots of a Protein-DNA Binding Landscape. \emph{Proc Natl
  Acad Sci U S A} \textbf{2010}, \emph{107}, 7751--6\relax
\mciteBstWouldAddEndPuncttrue
\mciteSetBstMidEndSepPunct{\mcitedefaultmidpunct}
{\mcitedefaultendpunct}{\mcitedefaultseppunct}\relax
\EndOfBibitem
\bibitem[Wolynes(1988)]{pmidPeterFugue}
Wolynes,~P.~G. Aperioidic Crystals: Biology, Chemistry and Physics in a Fugue
  with Stretto. \emph{AIP Conf. Proc.} \textbf{1988}, \emph{180}, 39--65\relax
\mciteBstWouldAddEndPuncttrue
\mciteSetBstMidEndSepPunct{\mcitedefaultmidpunct}
{\mcitedefaultendpunct}{\mcitedefaultseppunct}\relax
\EndOfBibitem
\bibitem[Wales(2012)]{pmid22615466}
Wales,~D.~J. Decoding the Energy Landscape: Extracting Structure, Dynamics and
  Thermodynamics. \emph{Philos Transact A Math Phys Eng Sci} \textbf{2012},
  \emph{370}, 2877--99\relax
\mciteBstWouldAddEndPuncttrue
\mciteSetBstMidEndSepPunct{\mcitedefaultmidpunct}
{\mcitedefaultendpunct}{\mcitedefaultseppunct}\relax
\EndOfBibitem
\bibitem[Denton et~al.(2002)Denton, Marshall, and Legge]{pmid12419661}
Denton,~M.~J.; Marshall,~C.~J.; Legge,~M. {{T}he Protein Folds as Platonic
  Forms: New Support for the Pre-{D}arwinian Conception of Evolution by Natural
  Law}. \emph{J. Theor. Biol.} \textbf{2002}, \emph{219}, 325--342\relax
\mciteBstWouldAddEndPuncttrue
\mciteSetBstMidEndSepPunct{\mcitedefaultmidpunct}
{\mcitedefaultendpunct}{\mcitedefaultseppunct}\relax
\EndOfBibitem
\bibitem[Schr{\"o}dinger(1944)]{whatislife}
Schr{\"o}dinger,~E. \emph{What Is Life?}; Cambridge University Press,
  1944\relax
\mciteBstWouldAddEndPuncttrue
\mciteSetBstMidEndSepPunct{\mcitedefaultmidpunct}
{\mcitedefaultendpunct}{\mcitedefaultseppunct}\relax
\EndOfBibitem
\bibitem[Schafer et~al.(2012)Schafer, Hoffman, Burger, Craig, Komives, and
  Wolynes]{pmid23251375}
Schafer,~N.~P.; Hoffman,~R.~M.; Burger,~A.; Craig,~P.~O.; Komives,~E.~A.;
  Wolynes,~P.~G. {{D}iscrete Kinetic Models from Funneled Energy Landscape
  Simulations}. \emph{PLoS ONE} \textbf{2012}, \emph{7}, e50635\relax
\mciteBstWouldAddEndPuncttrue
\mciteSetBstMidEndSepPunct{\mcitedefaultmidpunct}
{\mcitedefaultendpunct}{\mcitedefaultseppunct}\relax
\EndOfBibitem
\bibitem[Levy et~al.(2005)Levy, Cho, Shen, Onuchic, and Wolynes]{pmid15701699}
Levy,~Y.; Cho,~S.~S.; Shen,~T.; Onuchic,~J.~N.; Wolynes,~P.~G. {{S}ymmetry and
  Frustration in Protein Energy Landscapes: a near Degeneracy Resolves the
  {R}op Dimer-folding Mystery}. \emph{Proc. Natl. Acad. Sci. U.S.A.}
  \textbf{2005}, \emph{102}, 2373--2378\relax
\mciteBstWouldAddEndPuncttrue
\mciteSetBstMidEndSepPunct{\mcitedefaultmidpunct}
{\mcitedefaultendpunct}{\mcitedefaultseppunct}\relax
\EndOfBibitem
\bibitem[Hegler et~al.(2008)Hegler, Weinkam, and Wolynes]{pmid19436496}
Hegler,~J.~A.; Weinkam,~P.; Wolynes,~P.~G. {{T}he Spectrum of Biomolecular
  States and Motions}. \emph{HFSP J} \textbf{2008}, \emph{2}, 307--313\relax
\mciteBstWouldAddEndPuncttrue
\mciteSetBstMidEndSepPunct{\mcitedefaultmidpunct}
{\mcitedefaultendpunct}{\mcitedefaultseppunct}\relax
\EndOfBibitem
\bibitem[Monod et~al.(1965)Monod, Wyman, and Changeux]{pmid14343300}
Monod,~J.; Wyman,~J.; Changeux,~J.~P. On the Nature of Allosteric Transitions:
  a Plausible Model. \emph{J. Mol. Biol.} \textbf{1965}, \emph{12},
  88--118\relax
\mciteBstWouldAddEndPuncttrue
\mciteSetBstMidEndSepPunct{\mcitedefaultmidpunct}
{\mcitedefaultendpunct}{\mcitedefaultseppunct}\relax
\EndOfBibitem
\bibitem[Kuriyan and Eisenberg(2007)Kuriyan, and Eisenberg]{pmid18075577}
Kuriyan,~J.; Eisenberg,~D. {{T}he Origin of Protein Interactions and Allostery
  in Colocalization}. \emph{Nature} \textbf{2007}, \emph{450}, 983--990\relax
\mciteBstWouldAddEndPuncttrue
\mciteSetBstMidEndSepPunct{\mcitedefaultmidpunct}
{\mcitedefaultendpunct}{\mcitedefaultseppunct}\relax
\EndOfBibitem
\bibitem[Wang and Wolynes(2011)Wang, and Wolynes]{pmid21876141}
Wang,~S.; Wolynes,~P.~G. {{O}n the Spontaneous Collective Motion of Active
  Matter}. \emph{Proc. Natl. Acad. Sci. U.S.A.} \textbf{2011}, \emph{108},
  15184--15189\relax
\mciteBstWouldAddEndPuncttrue
\mciteSetBstMidEndSepPunct{\mcitedefaultmidpunct}
{\mcitedefaultendpunct}{\mcitedefaultseppunct}\relax
\EndOfBibitem
\bibitem[Jablonka and Raz(2009)Jablonka, and Raz]{pmid19606595}
Jablonka,~E.; Raz,~G. Transgenerational Epigenetic Inheritance: Prevalence,
  Mechanisms, and Implications for the Study of Heredity and Evolution. \emph{Q
  Rev Biol} \textbf{2009}, \emph{84}, 131--76\relax
\mciteBstWouldAddEndPuncttrue
\mciteSetBstMidEndSepPunct{\mcitedefaultmidpunct}
{\mcitedefaultendpunct}{\mcitedefaultseppunct}\relax
\EndOfBibitem
\bibitem[Pauling and Itano(1949)Pauling, and Itano]{pmid15395398}
Pauling,~L.; Itano,~H.~A. Sickle Cell Anemia a Molecular Disease.
  \emph{Science} \textbf{1949}, \emph{110}, 543--8\relax
\mciteBstWouldAddEndPuncttrue
\mciteSetBstMidEndSepPunct{\mcitedefaultmidpunct}
{\mcitedefaultendpunct}{\mcitedefaultseppunct}\relax
\EndOfBibitem
\bibitem[Treusch et~al.(2009)Treusch, Cyr, and Lindquist]{pmid19411847}
Treusch,~S.; Cyr,~D.~M.; Lindquist,~S. Amyloid Deposits: Protection Against
  Toxic Protein Species? \emph{Cell Cycle} \textbf{2009}, \emph{8},
  1668--74\relax
\mciteBstWouldAddEndPuncttrue
\mciteSetBstMidEndSepPunct{\mcitedefaultmidpunct}
{\mcitedefaultendpunct}{\mcitedefaultseppunct}\relax
\EndOfBibitem
\bibitem[Frauenfelder et~al.(1991)Frauenfelder, Sligar, and
  Wolynes]{pmid1749933}
Frauenfelder,~H.; Sligar,~S.~G.; Wolynes,~P.~G. {{T}he Energy Landscapes and
  Motions of Proteins}. \emph{Science} \textbf{1991}, \emph{254},
  1598--1603\relax
\mciteBstWouldAddEndPuncttrue
\mciteSetBstMidEndSepPunct{\mcitedefaultmidpunct}
{\mcitedefaultendpunct}{\mcitedefaultseppunct}\relax
\EndOfBibitem
\bibitem[Frauenfelder(2002)]{pmid11875198}
Frauenfelder,~H. Proteins: Paradigms of Complexity. \emph{Proc. Natl. Acad.
  Sci. U.S.A.} \textbf{2002}, \emph{99 Suppl 1}, 2479--80\relax
\mciteBstWouldAddEndPuncttrue
\mciteSetBstMidEndSepPunct{\mcitedefaultmidpunct}
{\mcitedefaultendpunct}{\mcitedefaultseppunct}\relax
\EndOfBibitem
\bibitem[Zhuravlev and Papoian(2010)Zhuravlev, and Papoian]{pmid20819242}
Zhuravlev,~P.~I.; Papoian,~G.~A. {{P}rotein Functional Landscapes, Dynamics,
  Allostery: a Tortuous Path towards a Universal Theoretical Framework}.
  \emph{Q. Rev. Biophys.} \textbf{2010}, \emph{43}, 295--332\relax
\mciteBstWouldAddEndPuncttrue
\mciteSetBstMidEndSepPunct{\mcitedefaultmidpunct}
{\mcitedefaultendpunct}{\mcitedefaultseppunct}\relax
\EndOfBibitem
\end{mcitethebibliography}

\end{document}